\documentclass[namedreferences]{solarphysics}

\usepackage[hyperref,optionalrh,natbib]{spr-sola-addons} 
\usepackage{graphicx}        
\usepackage{epstopdf}
\usepackage{hyperref}
\usepackage{color}           
\usepackage{enumitem}

\begin{document}
\begin{article}

\begin{opening}

\title{Oscillation Maps in the Broadband Radio Spectrum\\ of the 1 August 2010 Event}

\author{M. \surname{Karlick\'y}$^{1}$\sep
        J. \surname{Ryb\'ak}$^{2}$}

\runningauthor{M. Karlick\'y, J. Ryb\'ak}
\runningtitle{Oscillation Maps in Radio Spectra}

\institute{$^{1}$ Astronomical Institute, Academy of Sciences of
           the Czech Republic, 251 65 Ond\v{r}ejov, Czech Republic
           \email{karlicky@asu.cas.cz}\\
           $^{2}$ Astronomical Institute, Slovak Academy of Sciences,
           SK-05960 Tatransk\'a Lomnica, Slovak Republic
           \email{rybak@astro.sk}}

\begin{abstract}
We search for indications of waves in the 25\,--\,2000 MHz radio spectrum of the
1 August 2010 event (SOL2010-08-01T08:57:00L075C013), where fast propagating
waves in the solar corona with the periods of 181, 69, and 40 seconds were
detected in UV observations. Using the wavelet technique we construct a new
type of map of oscillations for selected periods in the whole domain of the
radio spectrum. While the oscillation with the period of 181 seconds was
recognized in the whole 25\,--\,2000 MHz radio spectrum, oscillations with
periods of 69 and 40 seconds were confirmed only in the 250\,--\,870 MHz
frequency range. In the 800\,--\,2000 MHz range we found periods of 50 and 80
seconds. Moreover, in the 250\,--\,870 MHz frequency range, the oscillation with
the period of about 420 seconds was detected. We also made maps of phases of
the 181-second oscillations in order to analyze their frequency drift. At the
beginning of the radio event, in the 2000\,--\,500 MHz frequency range the
phase of the 181-second oscillation drifts towards lower frequencies. On the
other hand, at frequencies 25\,--\,500 MHz we found that the phase is nearly
synchronous. While the phase drift at higher frequencies can be interpreted as
being caused by the UV wave, the synchronization of the phase on lower
frequencies is explained by the fast-electron beams, whose acceleration is
modulated by the UV wave. Owing to this modulation, the electron beams are
accelerated with the period of the UV wave (181 seconds). These beams propagate
upwards through the solar corona and generate the 25\,--\,500 MHz radio
emission with the 181-second period. Due to high beam velocity ($\approx$c/3,
where c is the light speed) the 25\,--\,500 MHz radio emission, corresponding
to a large interval of heights in the solar corona, is nearly synchronous.
\end{abstract}

\end{opening}

\section{Introduction}

In solar flares, oscillations and waves are commonly observed
in radio, X-ray, ultraviolet, and even in $\gamma$-ray emissions
~(\opencite{1984ApJ...279..857R,2003SoPh..218..183F,2005A&A...435..753W,2006A&A...452..343N,2010PPCF...52l4009N}).

The period of these oscillations ranges from sub-seconds to tens of minutes
~(\opencite{2006A&A...460..865M,2008SoPh..253..117T,2010SoPh..261..281K,2010SoPh..267..329K,
2011A&A...525A..88M,2014ApJ...791...44H,2014A&A...569A..12N}). Several
theoretical models explaining them have been proposed
~(\opencite{2009SSRv..149..119N}).

Various types of oscillations and modes of waves, especially in coronal loops, were
already studied numerically
(\opencite{2002ApJ...580L..85O,2003A&A...408..755D,2004A&A...414L..25N,
2005SSRv..121..115N,2005A&A...436..701S,2009EPJD...54..305J,2010ITPS...38.2243J,
2010A&A...521A..34K,2012ApJ...754..111O,2014ApJ...784..101P,2014ApJ...788...44M}).

Very interesting examples of the waves have been presented by
\cite{2011ApJ...736L..13L}. Based on the UV images observed by the {\it Atmospheric Imaging
Assembly} (AIA) onboard {\it Solar Dynamic Observatory} (SDO: \opencite{2012SoPh..275...17L})
during the 1 August 2010 flare, the authors showed that these waves are fast-mode
magnetosonic waves (having the periods of 181, 69, 40 seconds)
propagating with the speed of about 2200 km s$^{-1}$ upwards through the solar corona.

In the present article, for the same event on 1 August 2010, we search for any
indication of these waves in the radio spectrum. For this purpose, using the
wavelet technique, we constructed a new type of map of oscillations made from
the radio spectrum. The maps present power and phase of recognized
oscillations. Under the assumption that bursts in the radio spectra are
generated by the plasma-emission mechanism (in the frequency range considered
here this assumption is commonly accepted), such maps give us information about
the distribution of oscillations and waves in the vertical direction in the
gravitationally stratified solar atmosphere. A possible frequency drift of the
oscillations should express the vertical motion of these oscillations. An
attempt to make a similar map, but for sub-second (0.1 second) pulsations
during the 11 April 2001 flare, was already made by \cite{2010SoPh..261..281K},
but now we use a much more sophisticated and more general method.

The paper is structured as follows: In Sections 2 and 3 we present the
data and method. The results are summarized in Section 4, and their analysis and
interpretations are in Section 5. Conclusions are given in Section 6.

\section{Data}
The radio spectrum studied was observed during the 1 August 2010 flare ({\it
Hinode} Flare Catalog - event number 022480~(\opencite{2012SoPh..279..317W})).
This flare, classified as a GOES C3.2 flare, started at 07:25\,UT, peaked at
08:57\,UT, and ended at 10:25\,UT in NOAA active region 11092
(\opencite{2011ApJ...736L..13L,2010ApJ...725L..84L,2011JGRA..116.4108S}).

The radio spectrum consists of spectra from three radiospectrographs in
the time interval of 07:45 -- 09:00 UT:

\begin{itemize}
\item The \href{http://www.asu.cas.cz/~radio/}{{\it Ond\v{r}ejov radiospectrograph}} located at Ond\v{r}ejov, Czech Republic.
Radiospectrograph frequency range, frequency resolution, and time
resolution are 800\,--\,2000\,MHz, 4.7\,MHz, and 10\,ms, respectively
\citep{2008SoPh..253...95J}.

\item The \href{http://soleil.i4ds.ch/solarradio/}{{\it Phoenix-4 Bleien radiospectrograph}}
located at Bleien, Switzerland. Its frequency range, mean frequency
resolution, and time resolution are 175\,--\,870\,MHz, 3.61\,MHz, and
10\,ms, respectively.

\item The \href{http://www.izmiran.ru/stp/lars/}{{\it IZMIRAN radiospectrograph}}
located at Troitsk near Moscow, Russia. Its frequency range and time
resolution are 25\,--\,270\,MHz and 40\,ms, respectively. The frequency range
consists of four frequency sub-ranges with different frequency resolutions:
270\,--\,180\,MHz, 1.0\,MHz; 180\,--\,90\,MHz, 0.5\,MHz; 90\,--\,45\,MHz,
0.25\,MHz; and 45\,-\,25\,MHz, 0.14\,MHz.
\end{itemize}

For the analysis performed in this study, all of the data were re-sampled to a temporal
sampling convenient for calculations but still small enough to analyze the
desired periods of the radio signal: Ond\v{r}ejov and IZMIRAN
radiospectrographs to 2.0 seconds, and Phoenix-4 Bleien radiospectrograph to 2.5 seconds.
temporal sampling.

\begin{figure}[t]
\centerline{\hspace*{0.20\textwidth}
\includegraphics[bb= 0 0 512 512, width=1.0\textwidth, height=0.40\textheight]{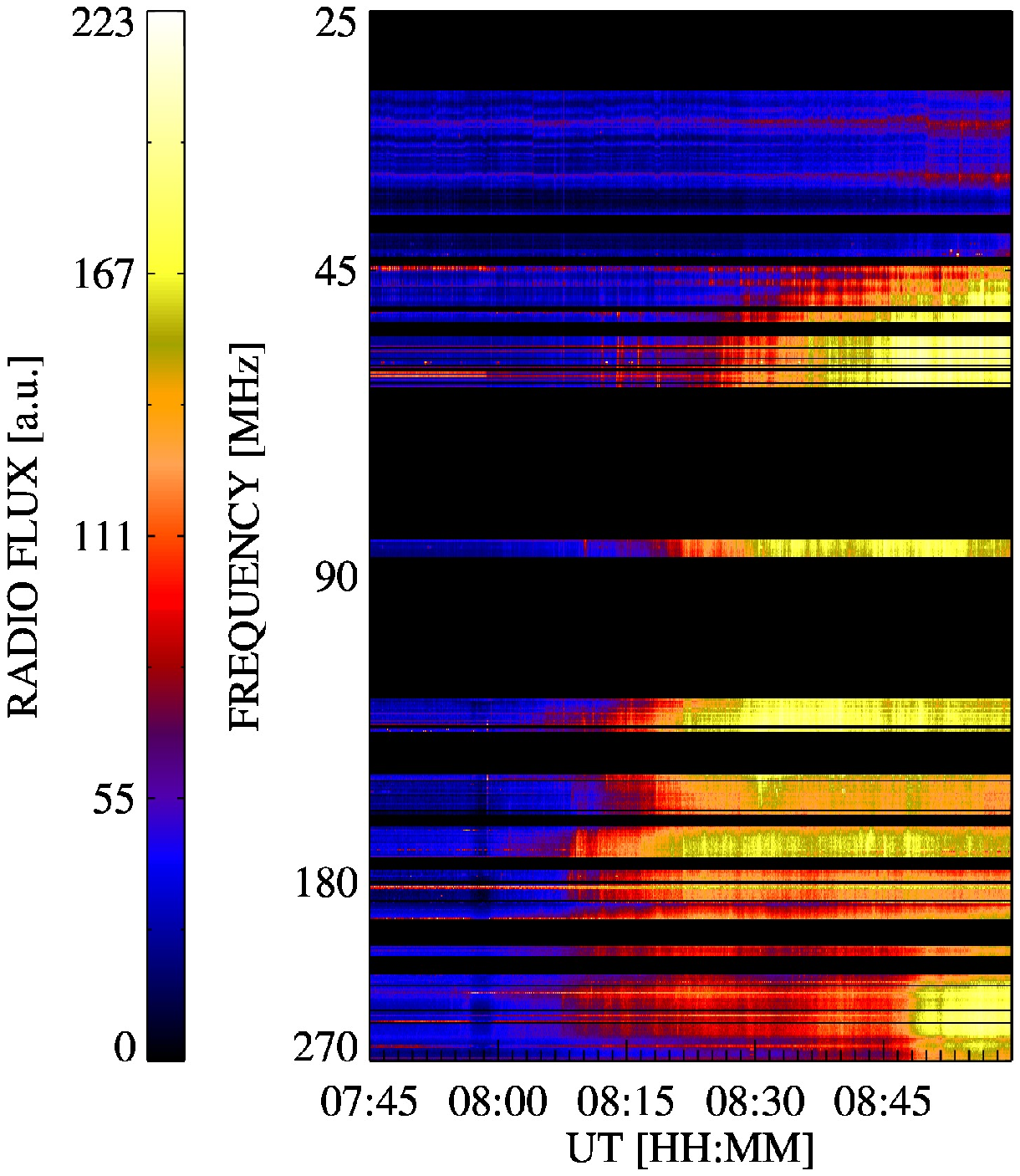}}
\vspace{-0.12\textwidth}
\centerline{\hspace*{0.20\textwidth}
\includegraphics[bb= 0 0 512 512, width=1.0\textwidth, height=0.40\textheight]{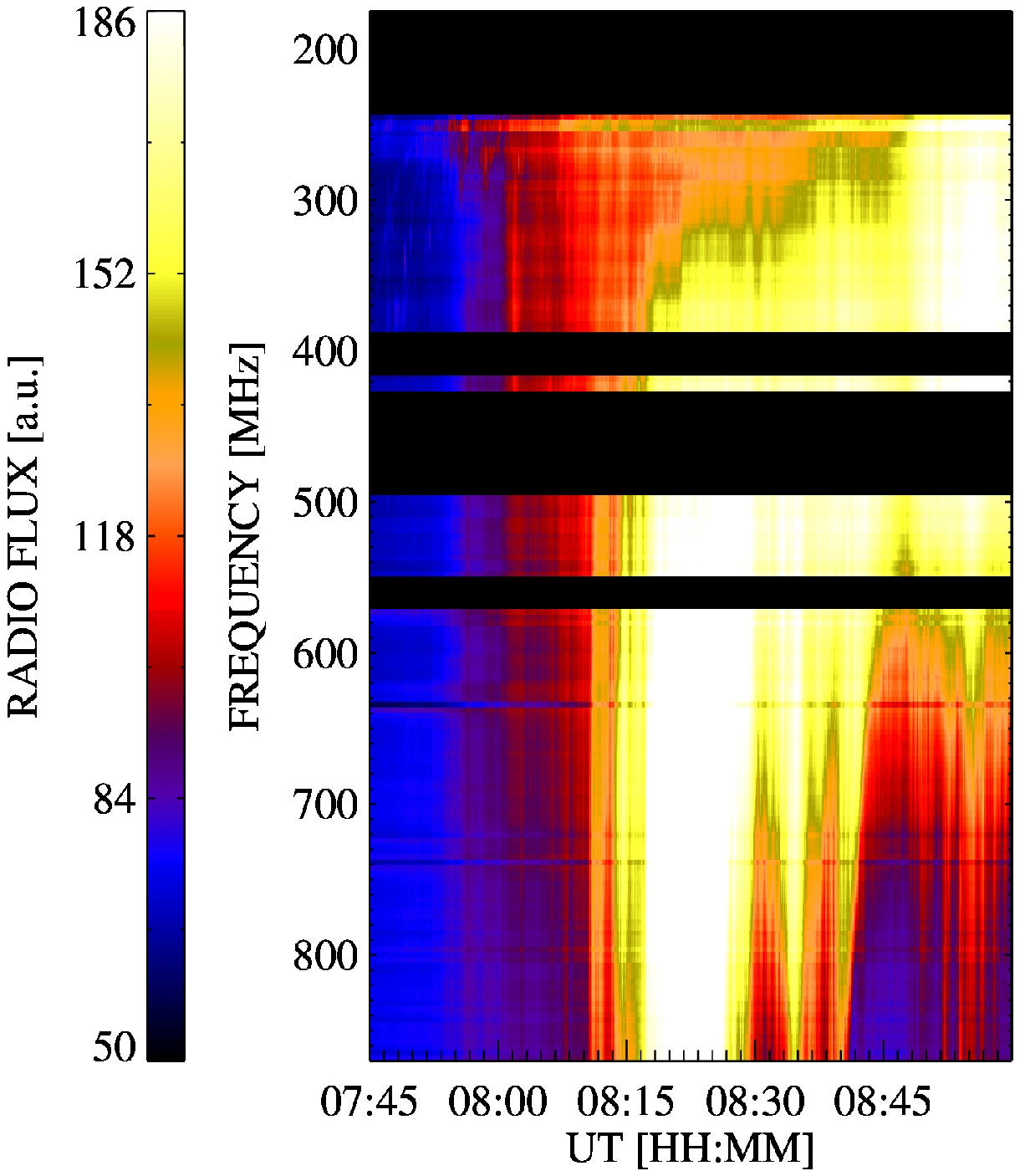}}
\vspace{-0.12\textwidth}
\centerline{\hspace*{0.20\textwidth}
\includegraphics[bb= 0 0 512 512, width=1.0\textwidth, height=0.40\textheight]{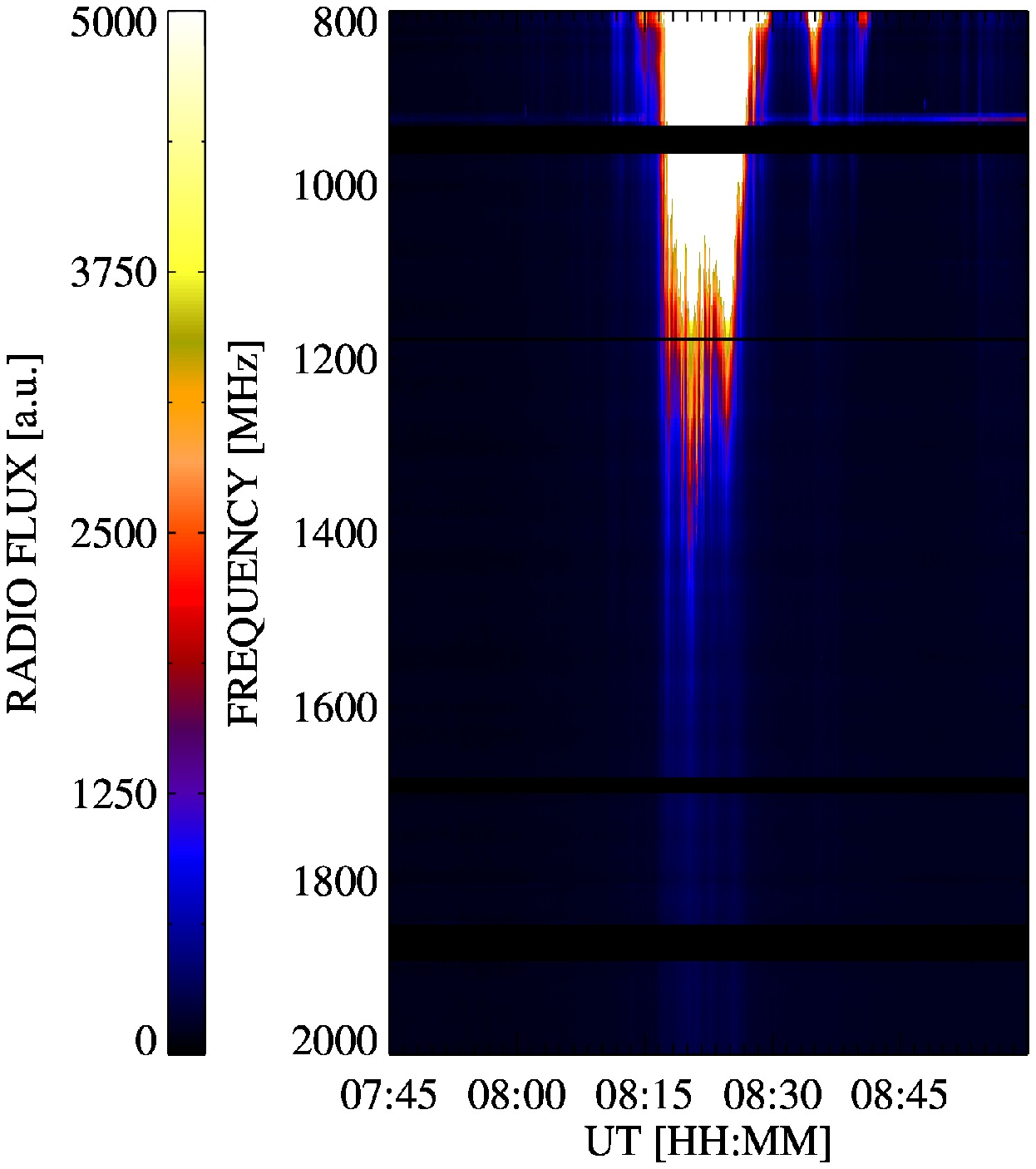}}
\caption{Radio spectra recorded by the
IZMIRAN (top), Phoenix-4 (middle), and Ond\v{r}ejov (bottom) radiospectrographs
for the common time interval 07:45\,--\,09:00\,UT on 1 August 2010.
Black-horizontal bands mark the data omitted from calculations due to presence of
radio noise.
White areas in the Phoenix-4 and Ond\v{r}ejov radio spectra indicate saturated
signal data.
}
\label{fig1}
\end{figure}

The broadband radio spectrum composed from the radio spectra of all the three
radiospectrographs is shown in Figure\,\ref{fig1}. Black bands show
frequencies, where the signal was heavily contaminated by artificial radio noise. These
frequencies were excluded from calculations. White areas indicate saturated
signals. The radio emission of this flare started at about 7:55 UT in
the 250\,--\,870 MHz range. Then during about 15 minutes the frequency range
of this emission increased up to about 2000\,--\,45 MHz interval. In the 250\,--\,2000
MHz range it consisted of many fast drifting pulses superimposed on the
continuum. The maximum of this burst (BM) slowly drifted from the frequency
$\approx$ 500 MHz at 8:22 UT to $\approx$ 350 MHz at 8:50 UT. In the 25\,--\,270 MHz range
the radio pulses were associated with pulsations and Type III bursts
superimposed on broadband continuum.

\begin{figure}[h]
\centerline{\hspace*{0.0\textwidth}
\includegraphics[bb= 0 0 440 440, clip, width=0.55\textwidth, height=0.35\textheight]{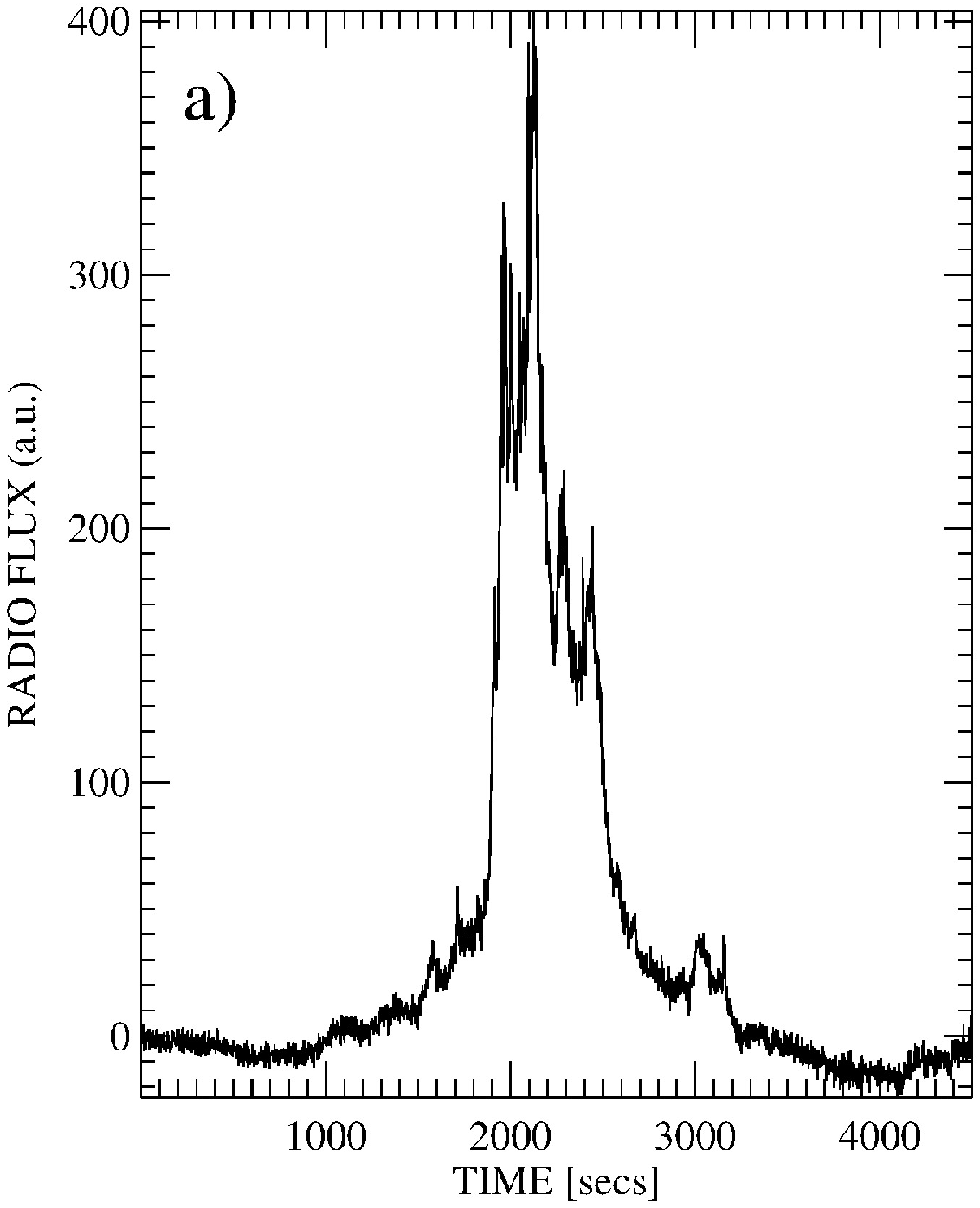}
\hspace*{-0.05\textwidth}
\includegraphics[bb= 0 0 440 440, clip, width=0.55\textwidth, height=0.35\textheight]{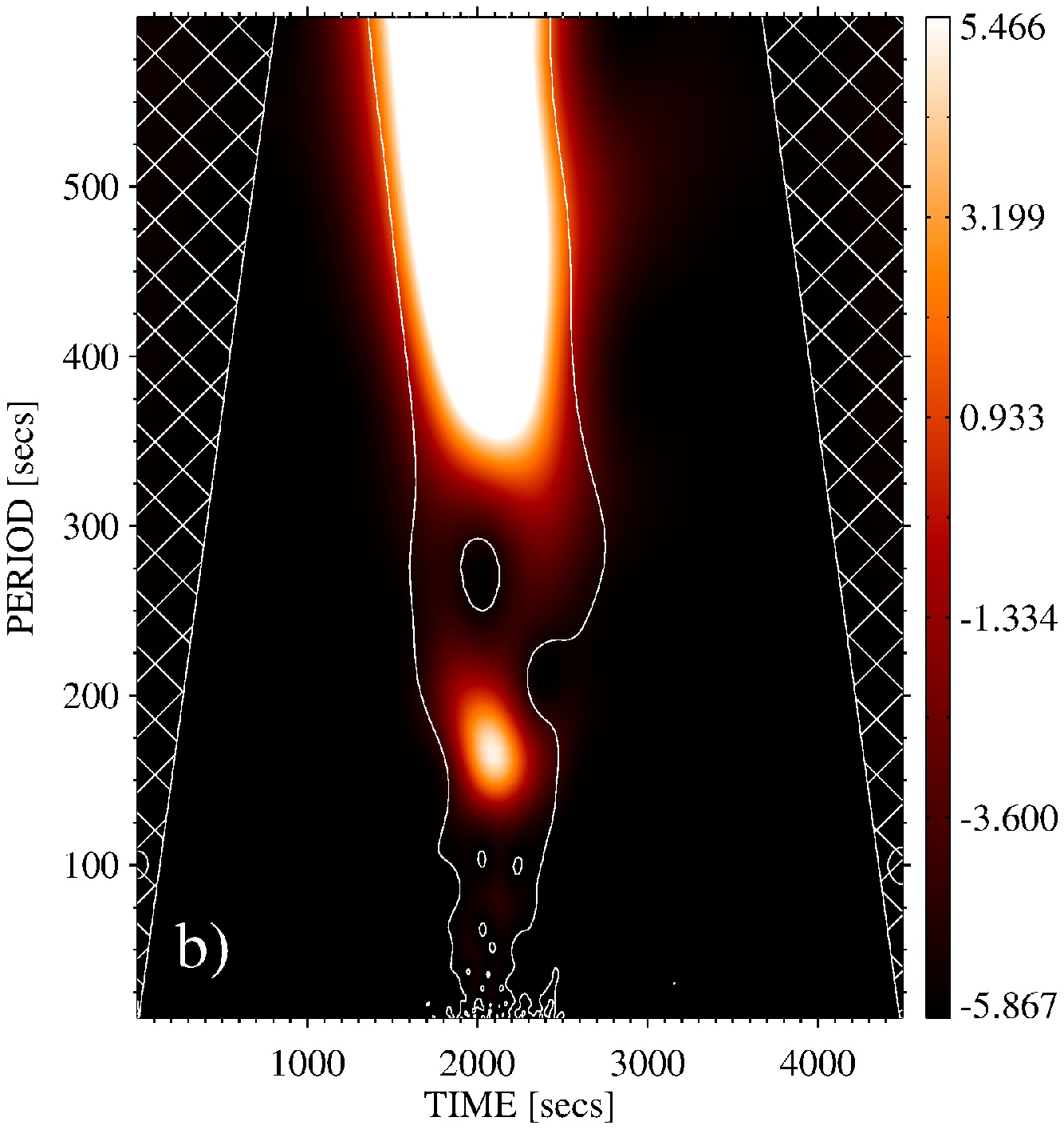}}
\vspace{-0.02\textwidth}
\centerline{\hspace*{0.0\textwidth}
\includegraphics[bb= 0 0 440 440, clip, width=0.55\textwidth, height=0.35\textheight]{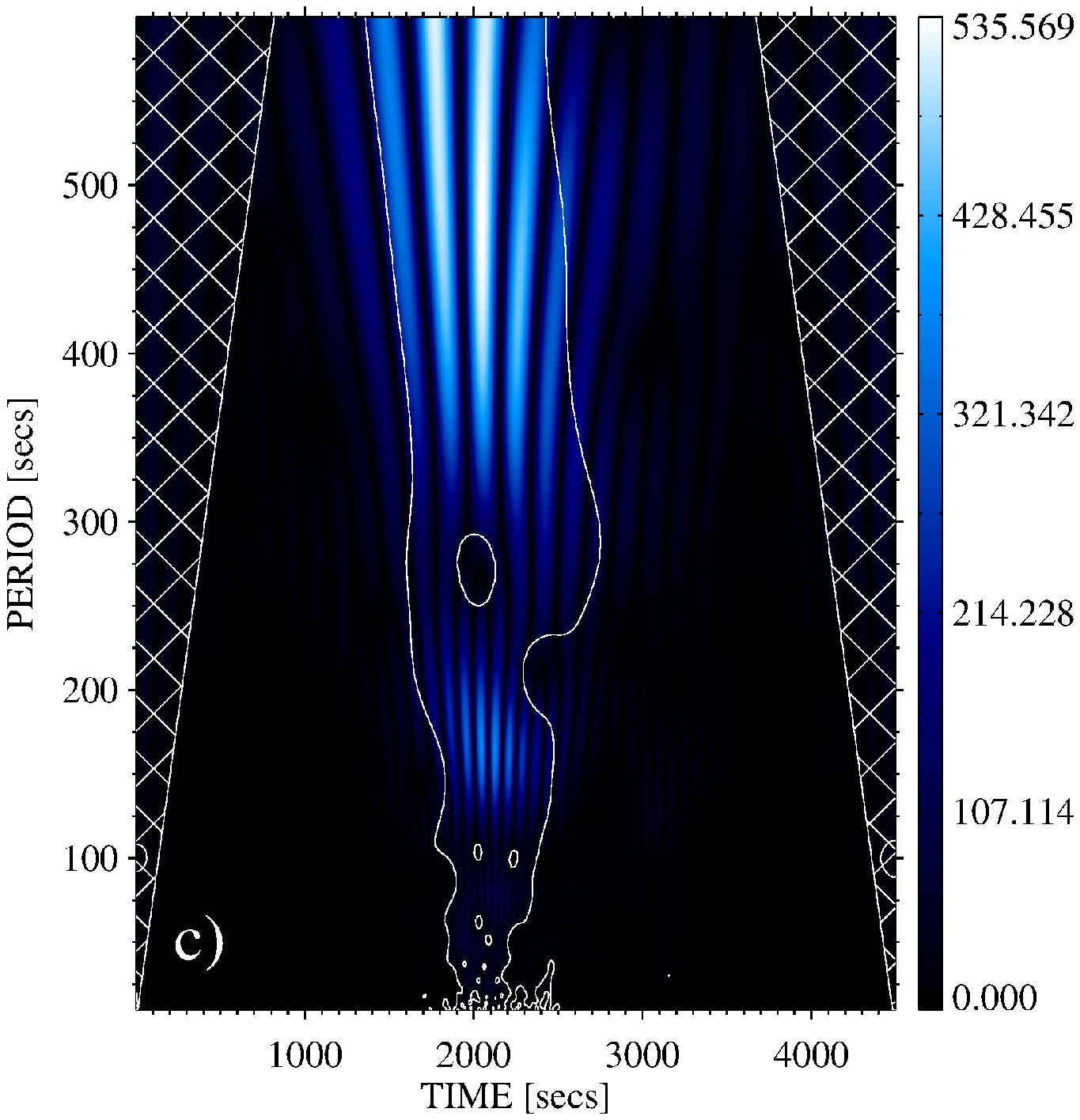}
\hspace*{-0.05\textwidth}
\includegraphics[bb= 0 0 440 440, clip, width=0.55\textwidth, height=0.35\textheight]{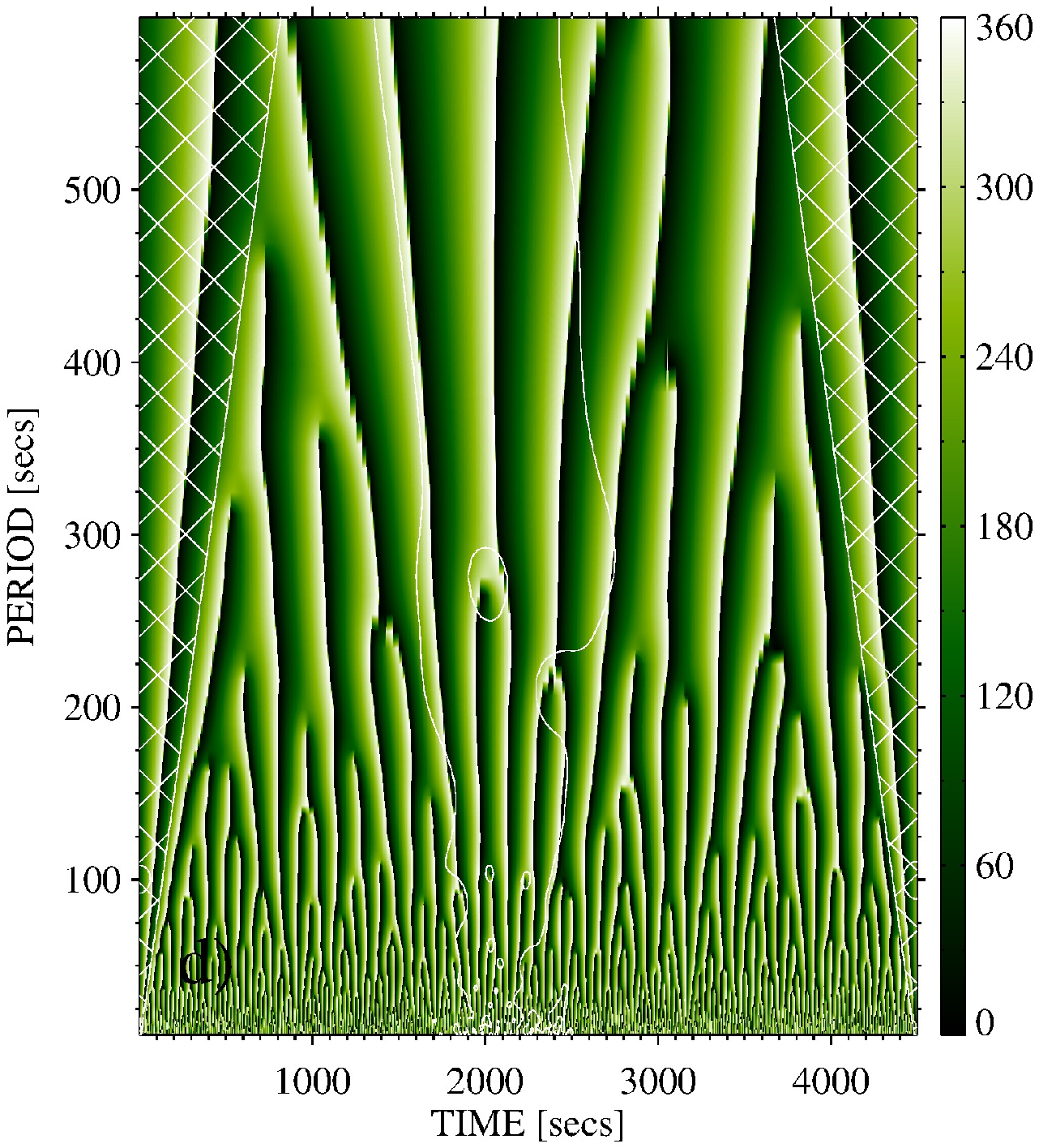}}
\caption{Example of the analysis of the radio-signal time series taken from the Ond\v{r}ejov radio spectrum
at the frequency 1600 MHz: a) radio flux {\it vs.} time with a background subtracted, b) WT power,
c) WT amplitude, and d) WT phase.
White contours in b) and c) mark results above the selected significance level and white
cross-hatched areas designate areas outside the cone of influence.
}
\label{fig2}
\end{figure}

\begin{figure}[t]
\centerline{\hspace*{0.0\textwidth}
\includegraphics[bb= 0 0 425 425, width=0.85\textwidth, height=0.35\textheight]{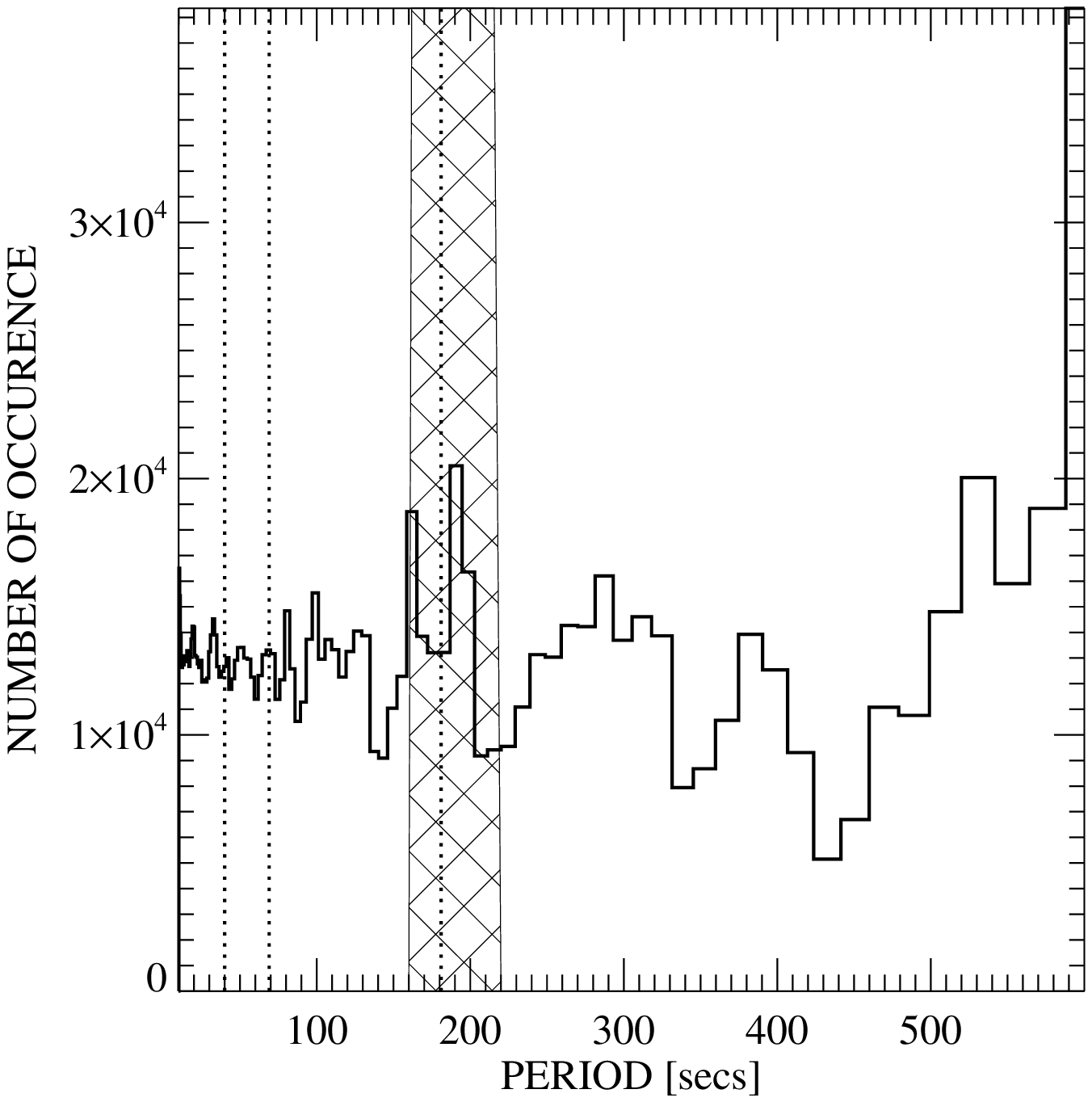}}
\centerline{\hspace*{0.0\textwidth}
\includegraphics[bb= 0 0 425 425, width=0.85\textwidth, height=0.35\textheight]{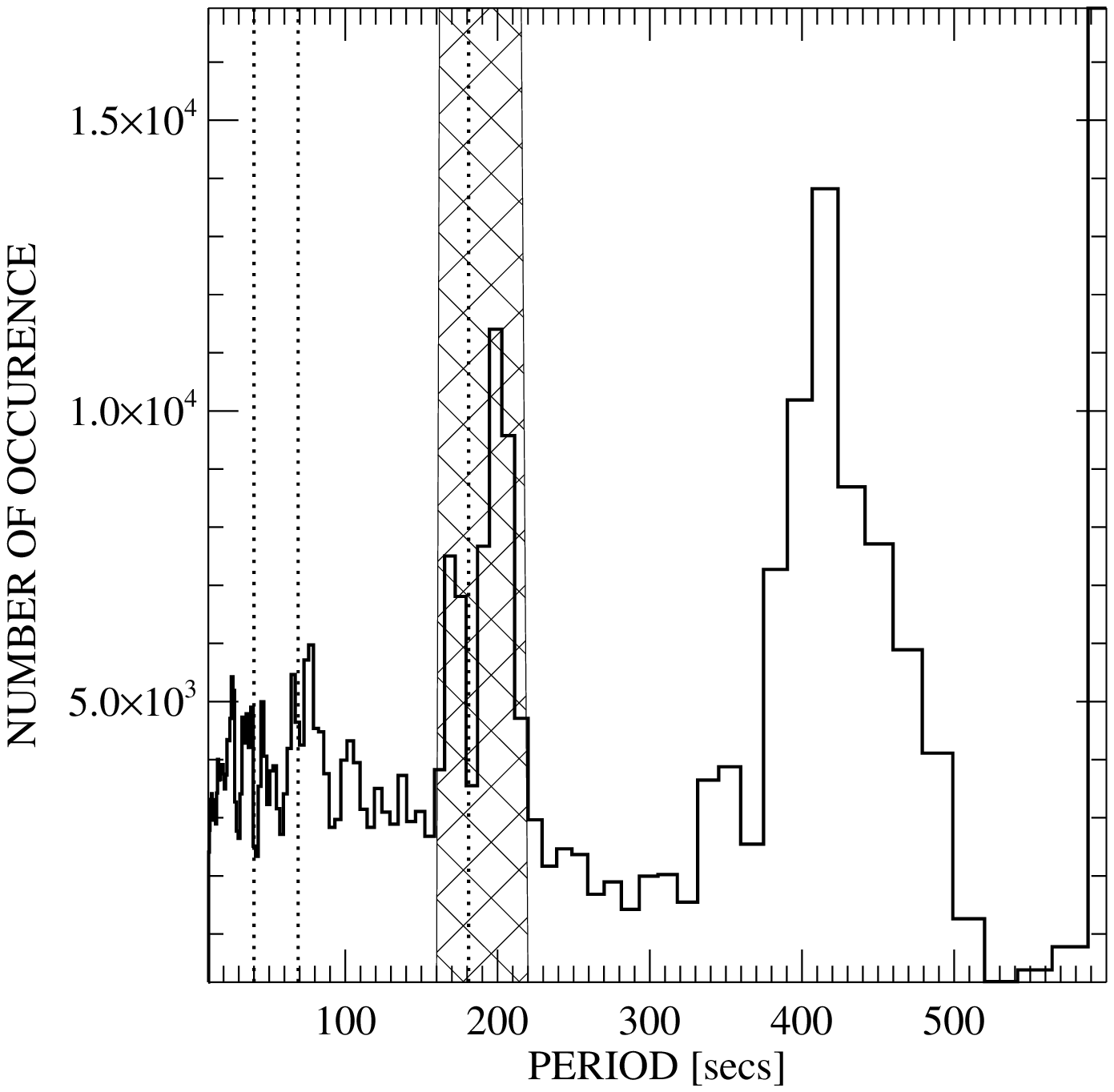}}
\centerline{\hspace*{0.0\textwidth}
\includegraphics[bb= 0 0 425 425, width=0.85\textwidth, height=0.35\textheight]{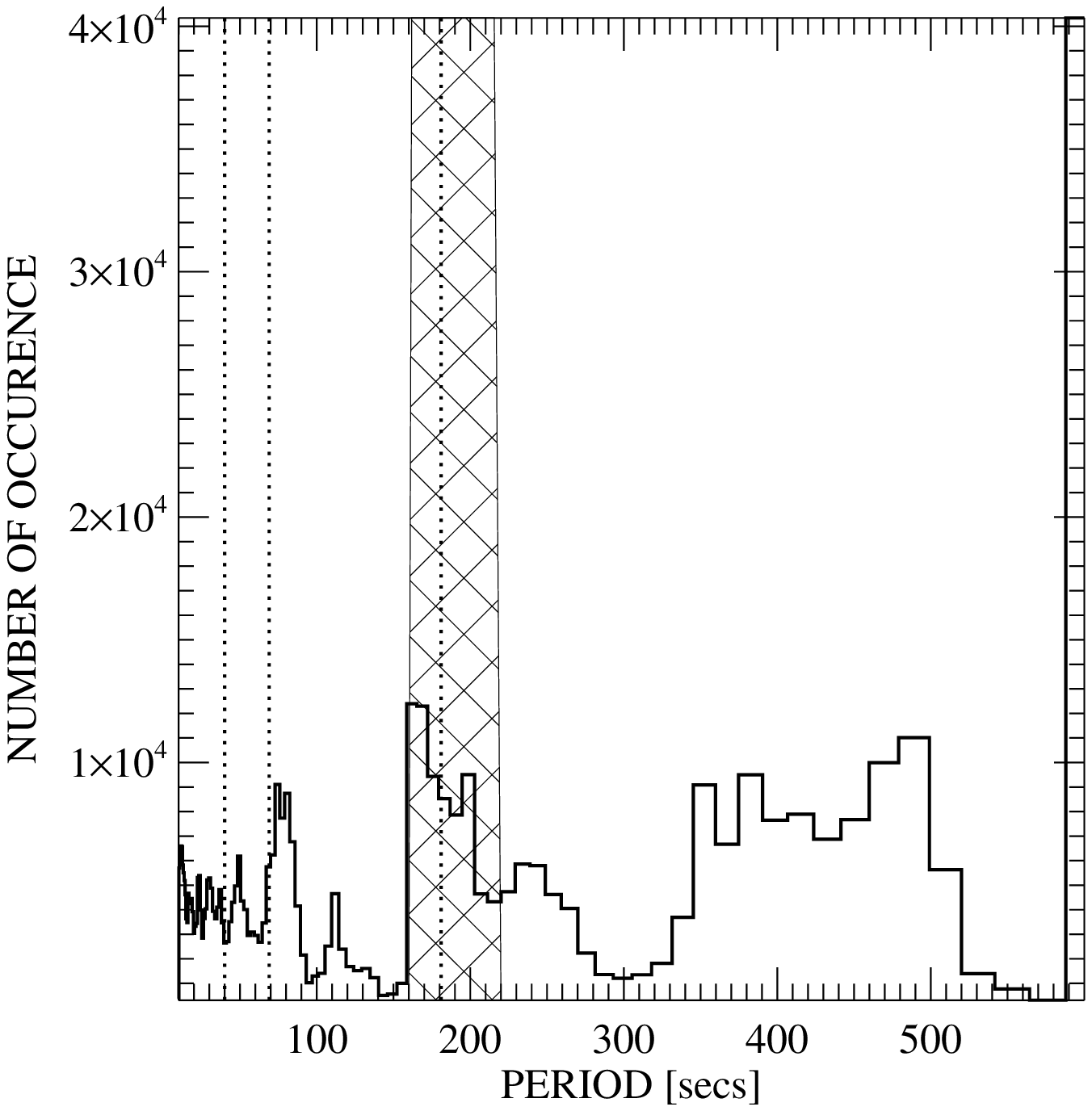}}
\caption{Histograms of the significant WT periods determined by the wavelet transformation of the radio-flux time series
at all frequencies of the radio spectra (IZMIRAN (top),  Phoenix-4 (middle), and Ond\v{r}ejov (bottom)).
The vertical-dotted lines show the periods found in UV observations (40, 69, and 181 seconds).
A cross-hatched band marks the interval of periods around 181 seconds,
considered for further detailed analysis (see Figures~\ref{fig4},~\ref{fig5}, and~\ref{fig6}).}
\label{fig3}
\end{figure}

\begin{figure}[h]
\centerline{\hspace*{0.1\textwidth}
\includegraphics[bb= 0 0 512 512, width=1.0\textwidth, height=0.4\textheight]{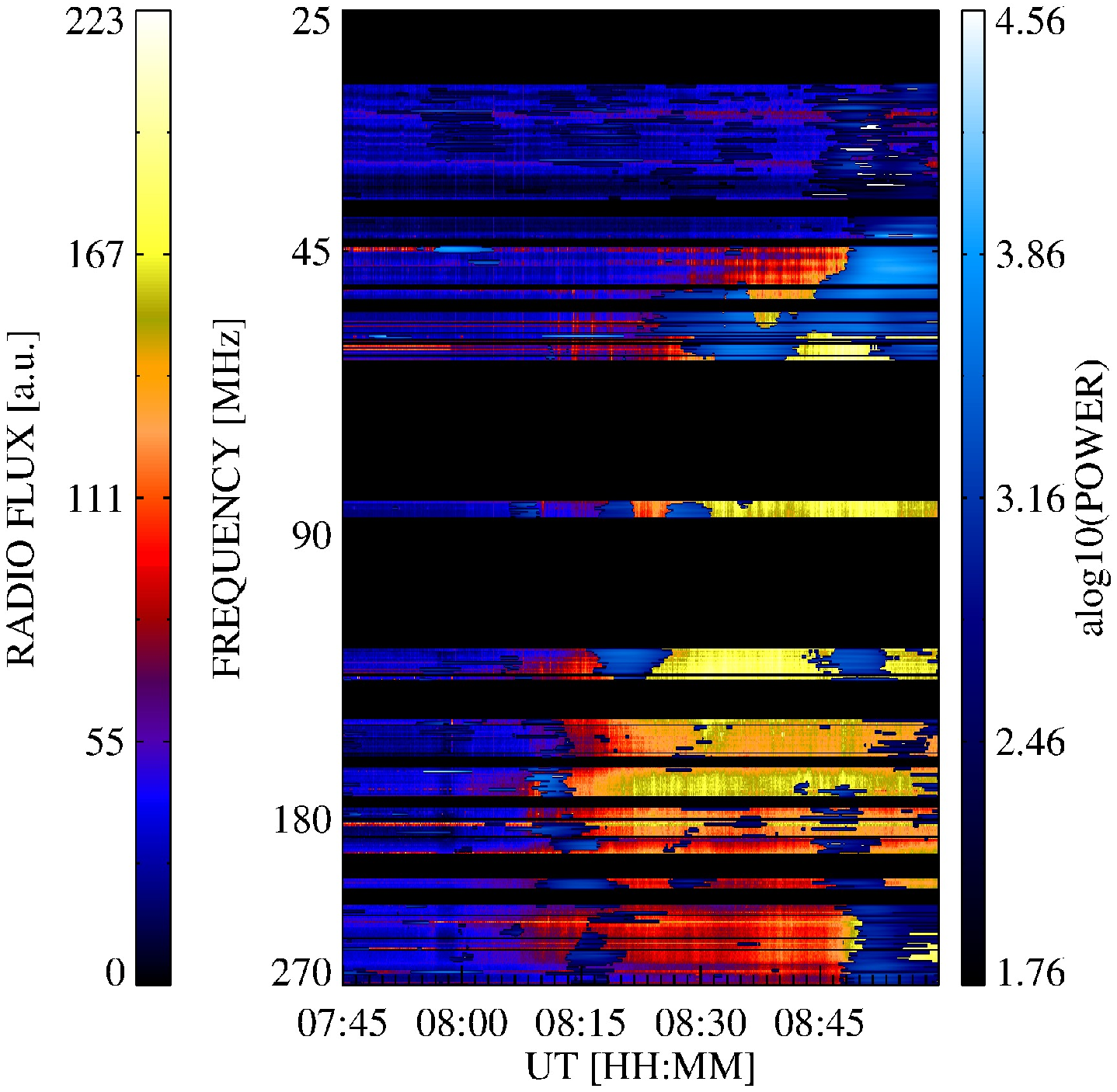}}
\vspace{-0.10\textwidth}
\centerline{\hspace*{0.1\textwidth}
\includegraphics[bb= 0 0 512 512, width=1.0\textwidth, height=0.4\textheight]{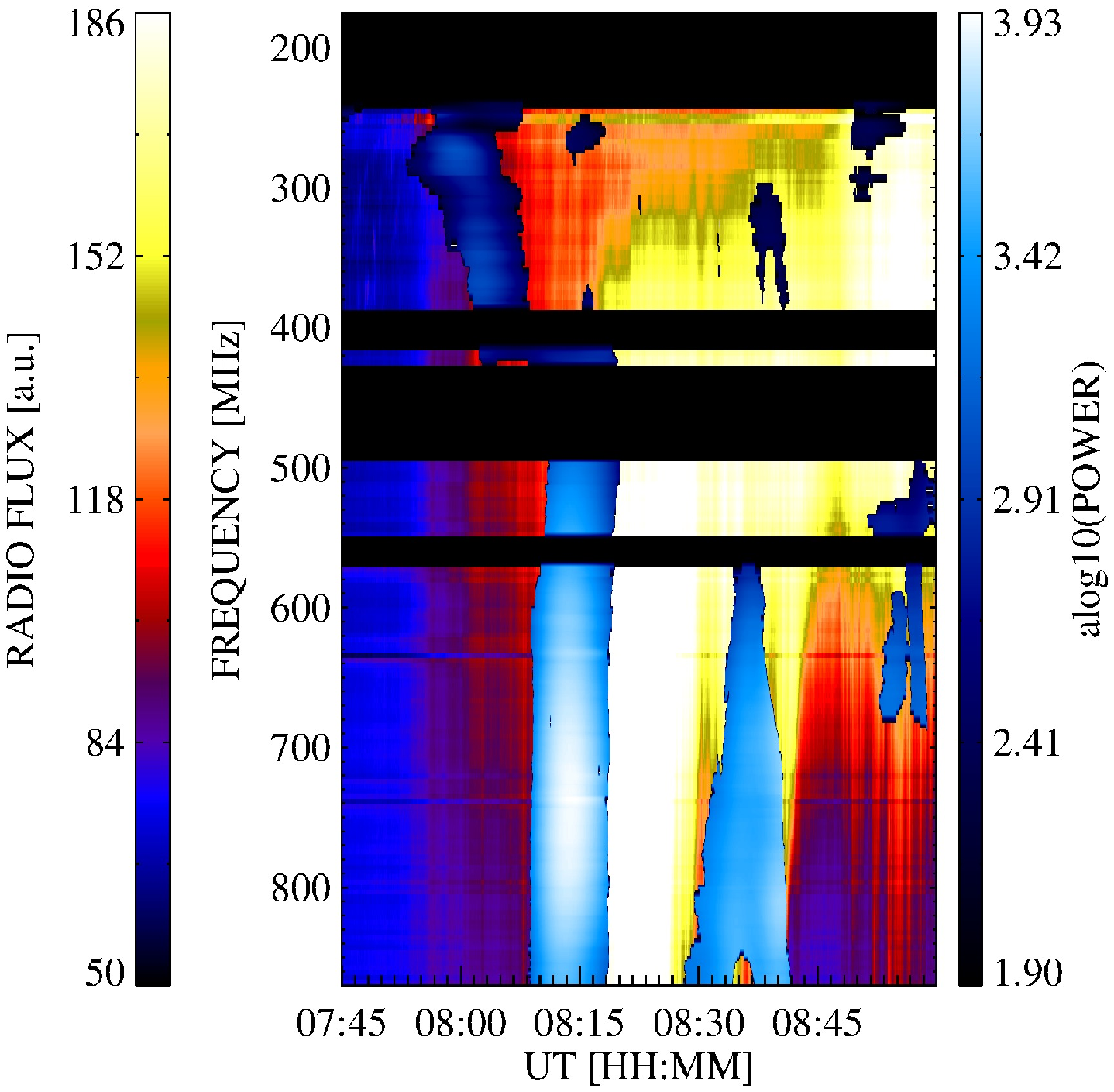}}
\vspace{-0.10\textwidth}
\centerline{\hspace*{0.1\textwidth}
\includegraphics[bb= 0 0 512 512, width=1.0\textwidth, height=0.4\textheight]{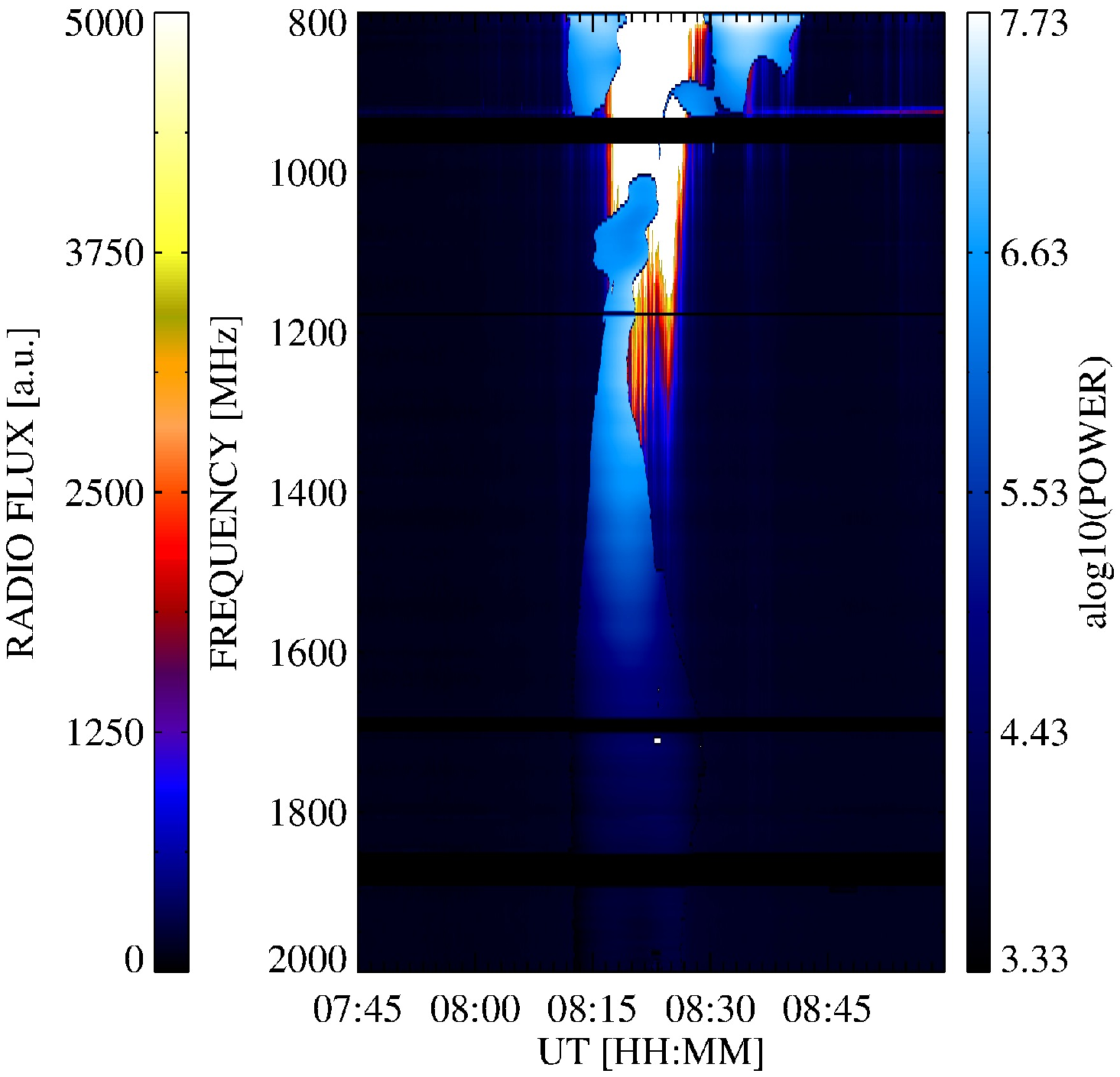}}
\caption{Radio spectra recorded by the
IZMIRAN ({\it top}), Phoenix-4 (middle), and Ond\v{r}ejov (bottom) radiospectrographs
over-plotted by the light-blue areas, where the significant WT power for the period interval around 181 seconds (see the cross-hatched band
in Figure~\ref{fig3}) was found.}
\label{fig4}
\end{figure}

\begin{figure}[h]
\centerline{\hspace*{0.1\textwidth}
\includegraphics[bb= 0 0 512 512, width=1.0\textwidth, height=0.4\textheight]{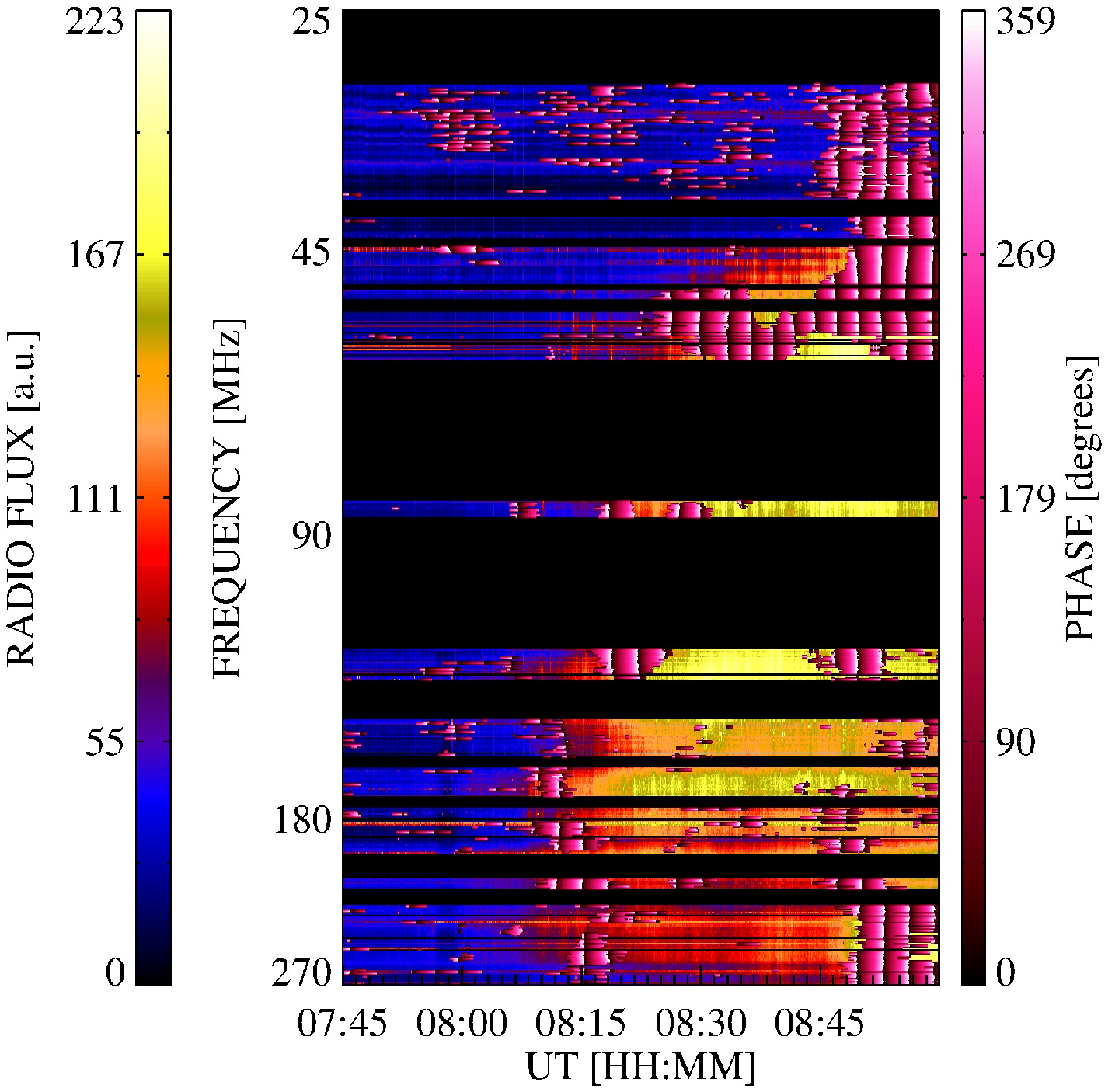}}
\vspace{-0.10\textwidth}
\centerline{\hspace*{0.1\textwidth}
\includegraphics[bb= 0 0 512 512, width=1.0\textwidth, height=0.4\textheight]{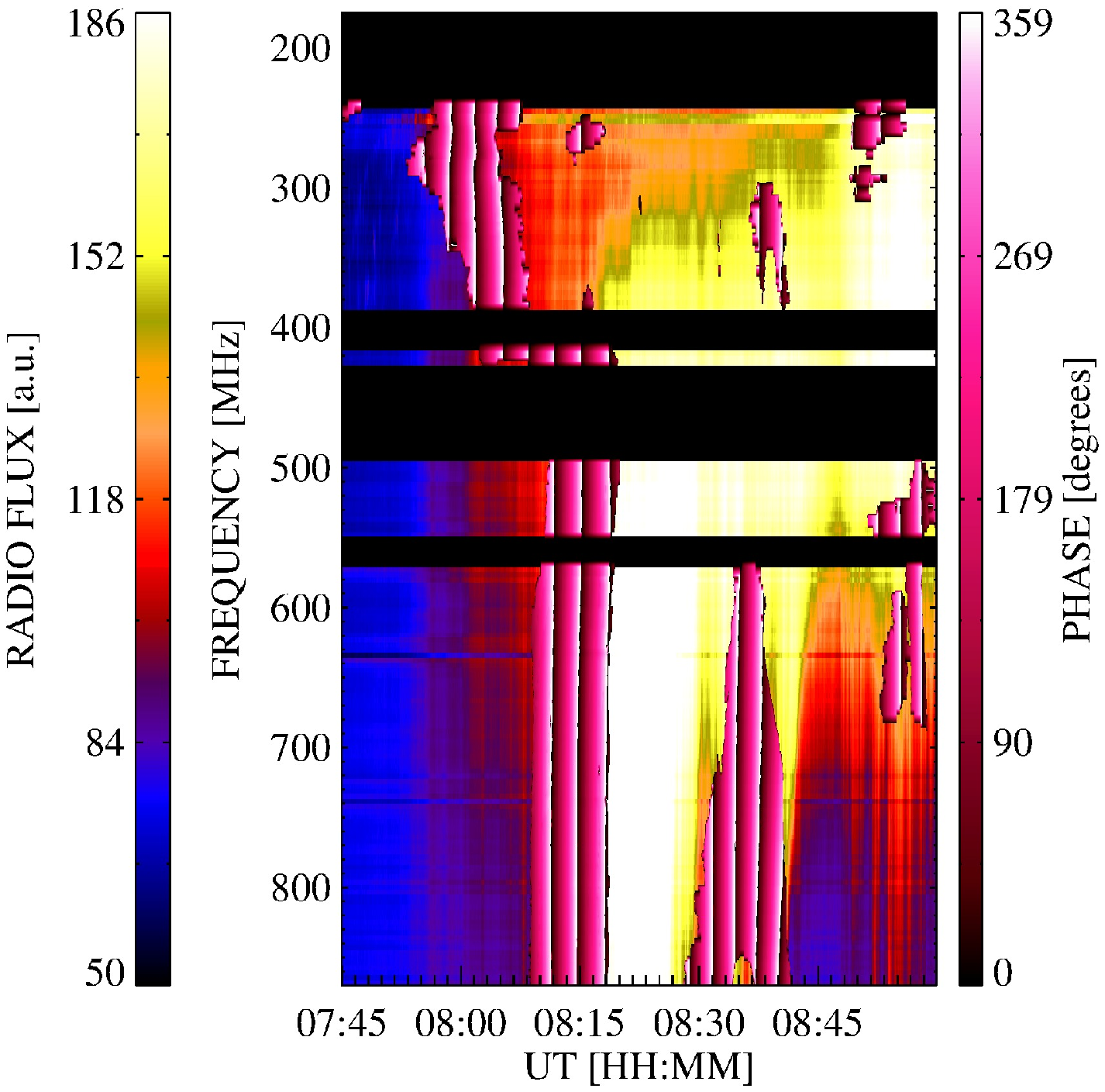}}
\vspace{-0.10\textwidth}
\centerline{\hspace*{0.1\textwidth}
\includegraphics[bb= 0 0 512 512, width=1.0\textwidth, height=0.4\textheight]{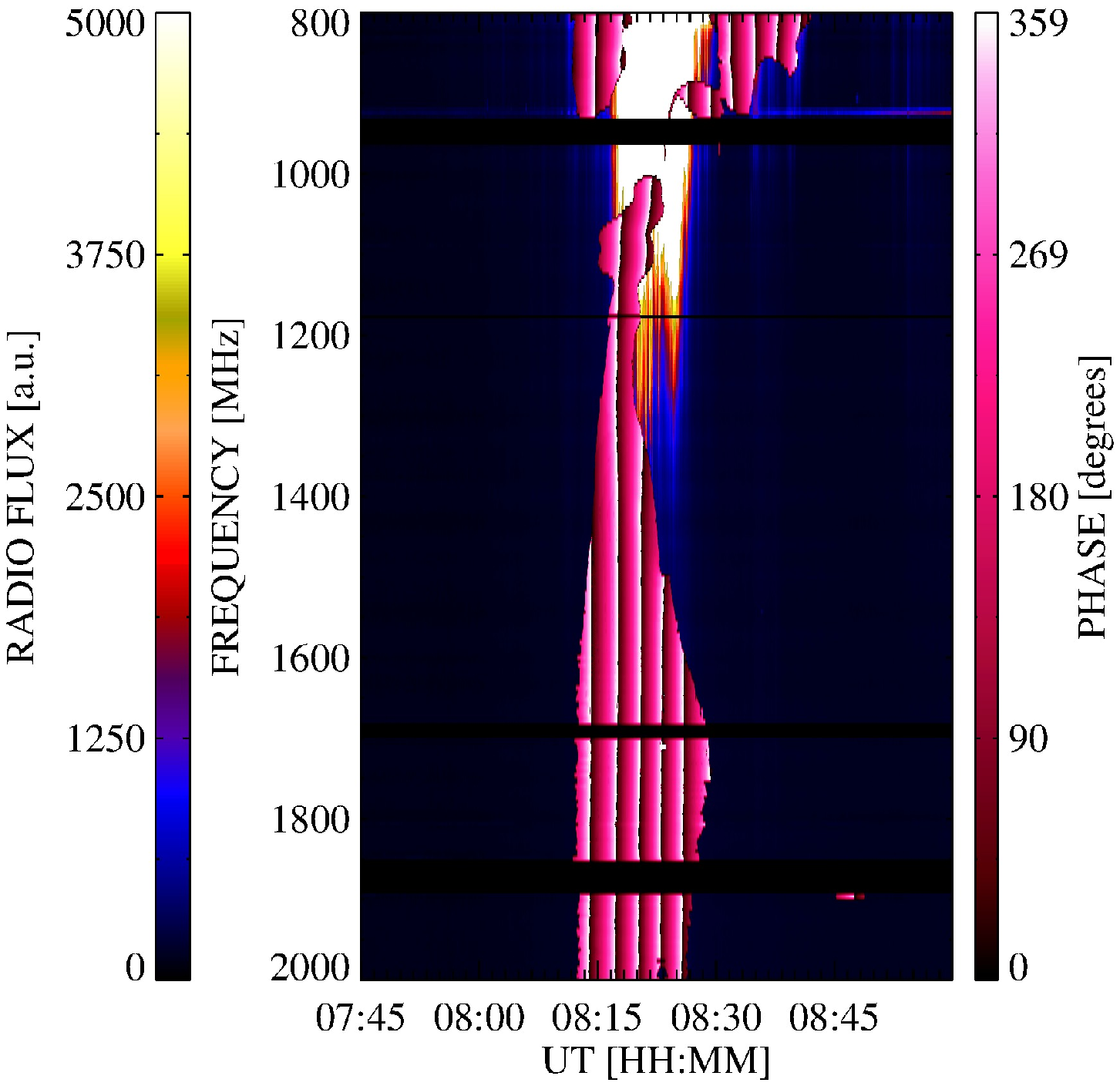}}
\caption{Radio spectra recorded by the
IZMIRAN (top), Phoenix-4 (middle), and Ond\v{r}ejov (bottom) radiospectrographs
over-plotted by the areas as in Figure~\ref{fig4}, but with the pink bands, showing the wave phase.
The black borders of these bands indicate zero phase.}
\label{fig5}
\end{figure}

\begin{figure}[h]
\centerline{\hspace*{0.1\textwidth}
\includegraphics[bb= 0 0 512 512, width=1.0\textwidth, height=0.4\textheight]{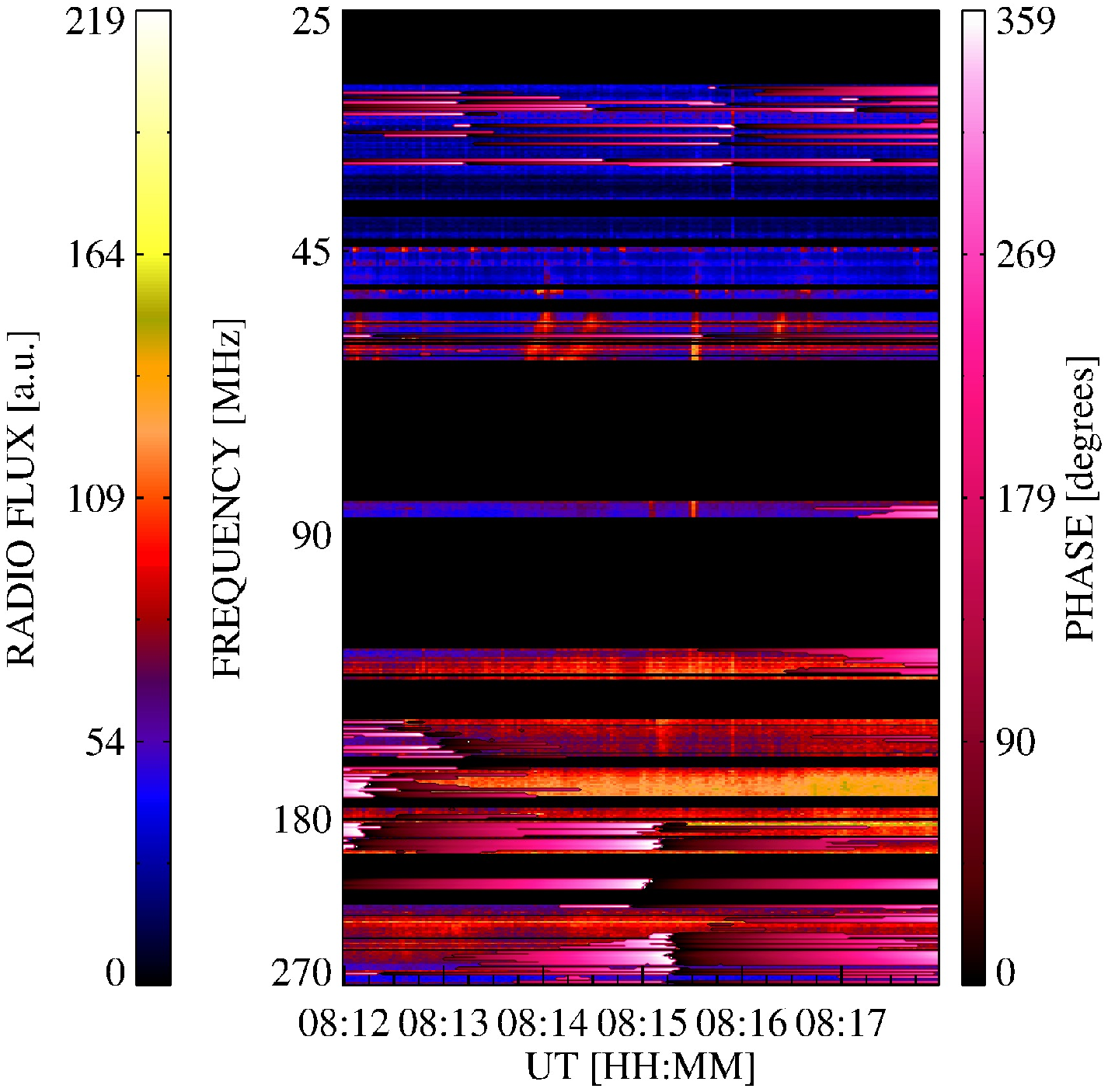}}
\vspace{-0.08\textwidth}
\centerline{\hspace*{0.1\textwidth}
\includegraphics[bb= 0 0 512 512, width=1.0\textwidth, height=0.4\textheight]{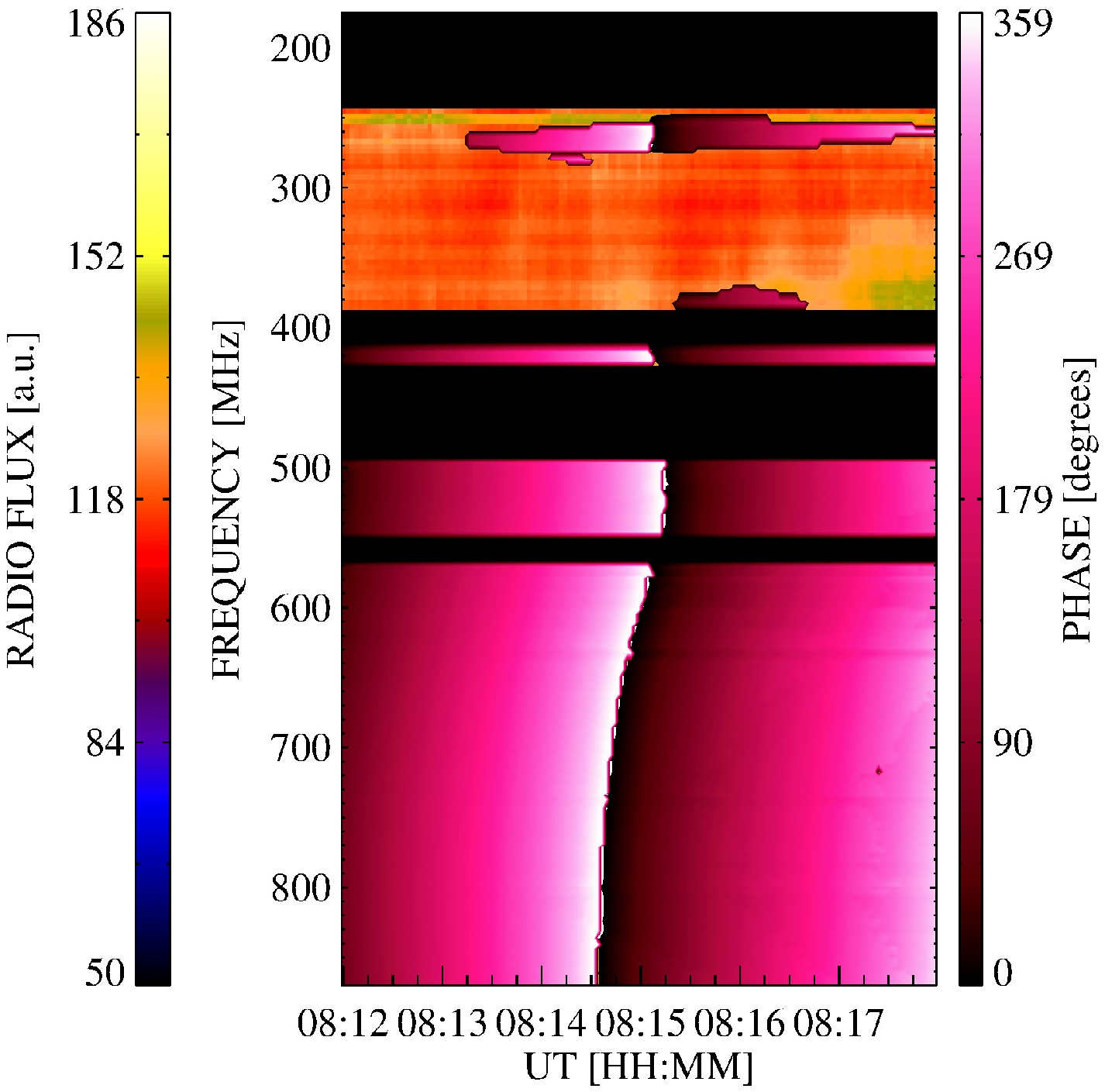}}
\vspace{-0.08\textwidth}
\centerline{\hspace*{0.1\textwidth}
\includegraphics[bb= 0 0 512 512, width=1.0\textwidth, height=0.4\textheight]{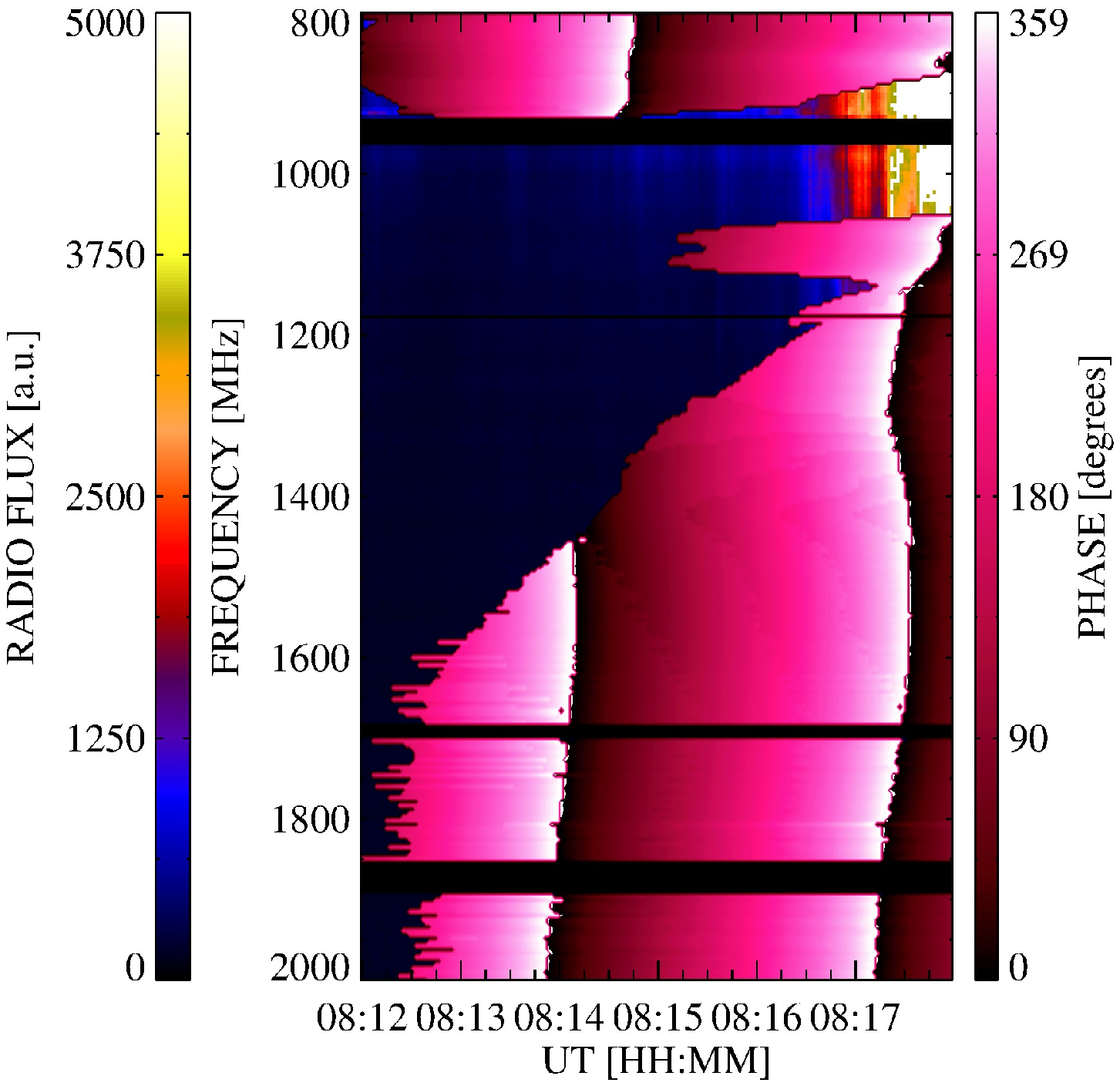}
}
\caption{Radio spectra recorded by the IZMIRAN (top), Phoenix-4 (middle), and Ond\v{r}ejov
(bottom) radiospectrographs over-plotted by the areas of the WT phase (the pink bands) but
just for the temporal interval 08:12\,--\,08:18\,UT.
The black borders of these bands indicate zero phase.}
\label{fig6}
\end{figure}


\begin{figure}[h]
\centerline{\hspace*{0.20\textwidth}
\includegraphics[bb= 0 0 512 512, width=1.0\textwidth, height=0.4\textheight]{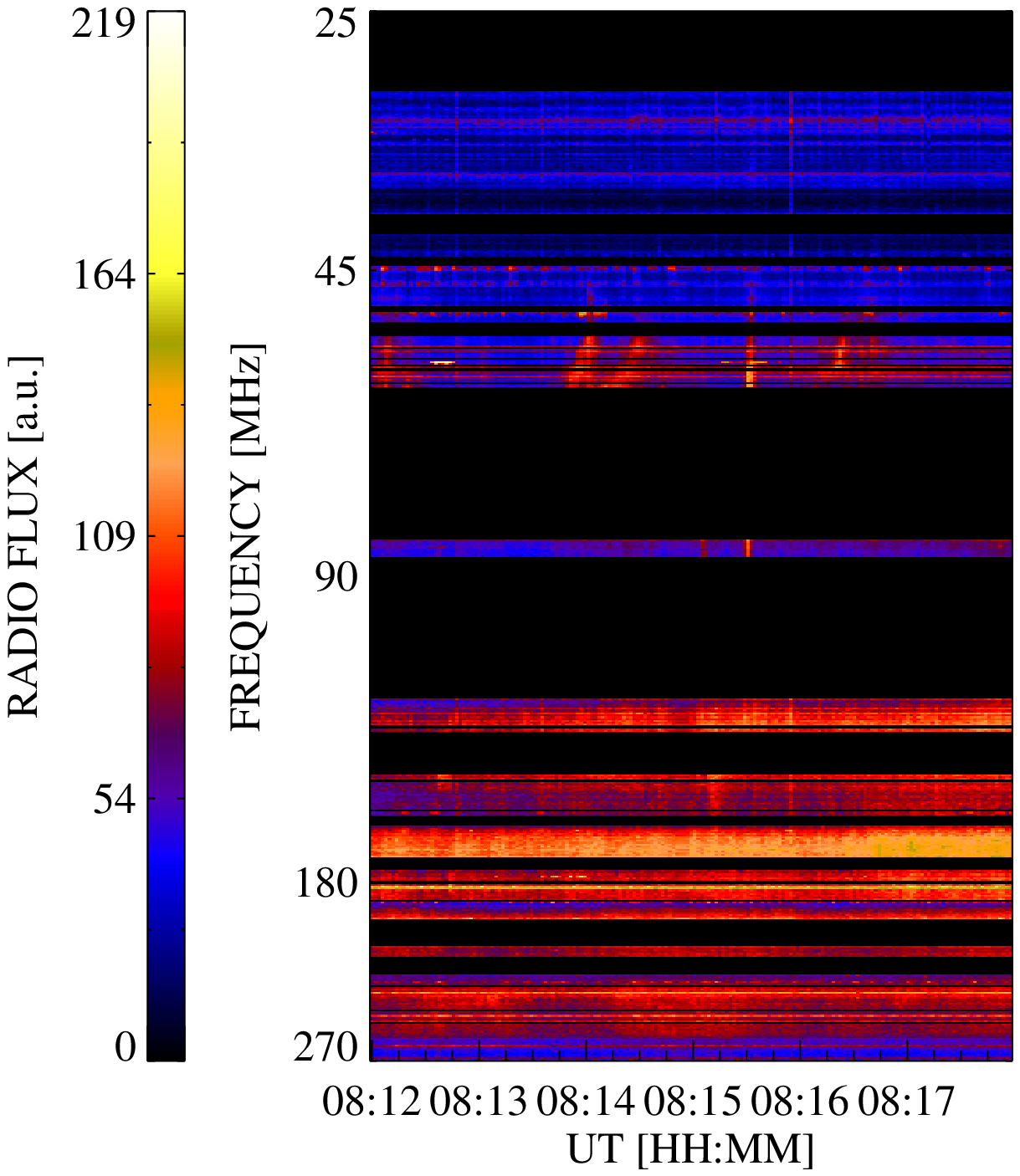}}
\vspace{-0.12\textwidth}
\centerline{\hspace*{0.20\textwidth}
\includegraphics[bb= 0 0 512 512, width=1.0\textwidth, height=0.4\textheight]{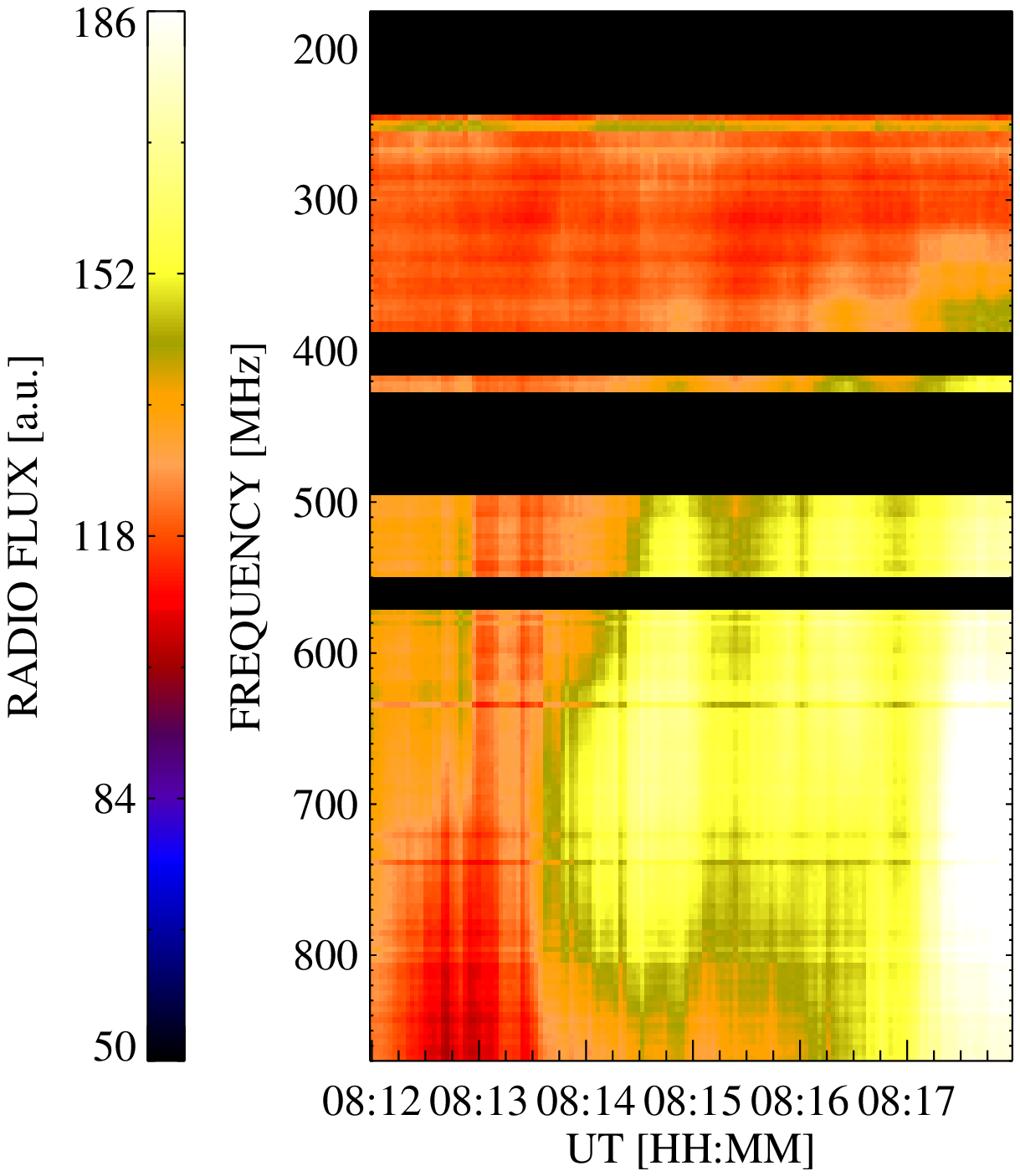}}
\vspace{-0.12\textwidth}
\centerline{\hspace*{0.20\textwidth}
\includegraphics[bb= 0 0 512 512, width=1.0\textwidth, height=0.4\textheight]{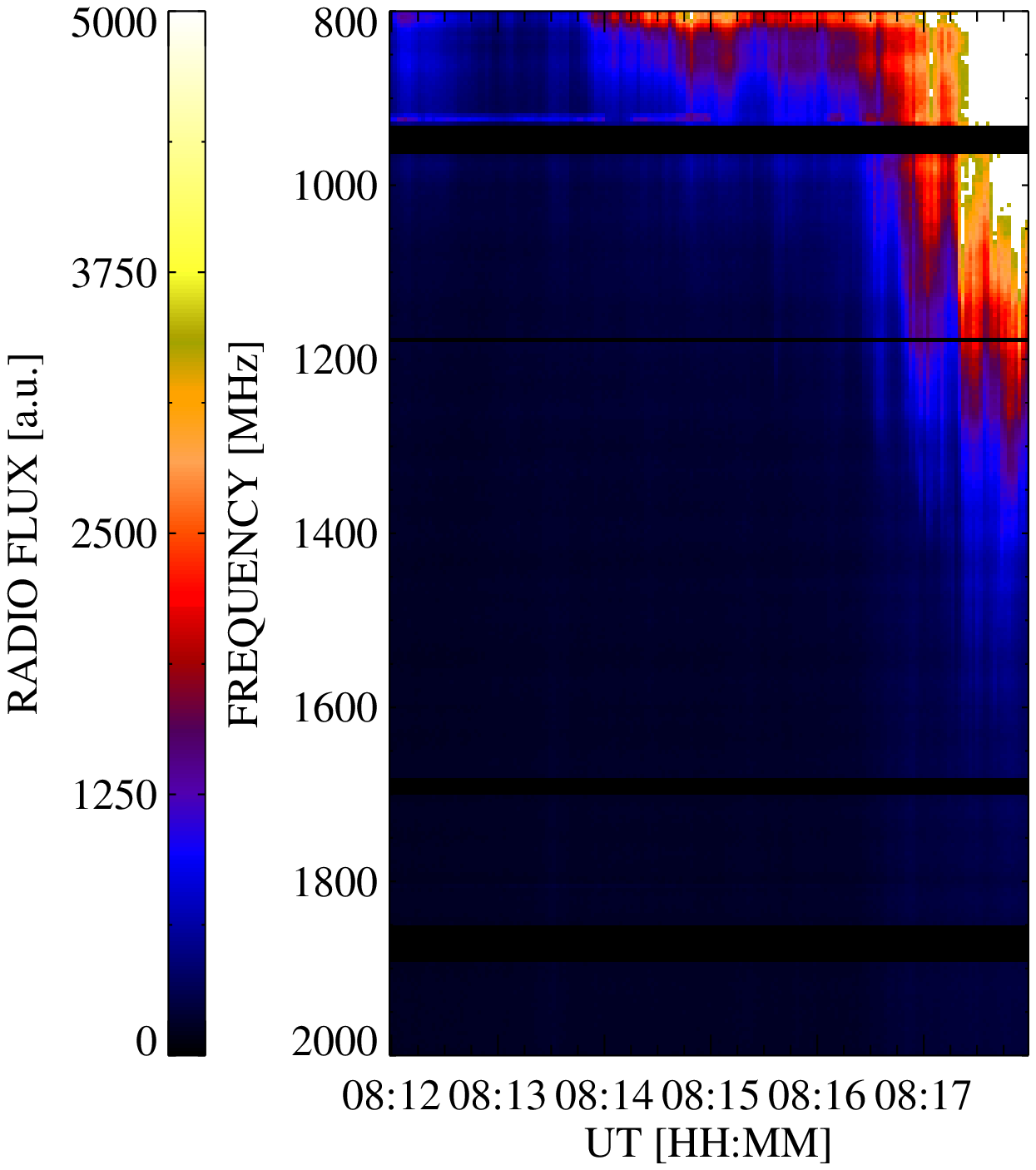}}
\caption{Original radio spectra recorded by the IZMIRAN (top), Phoenix-4 (middle),
and Ond\v{r}ejov (bottom) radiospectrographs without any overplot but
just for the temporal interval 08:12\,--\,08:18\,UT.}
\label{fig7}
\end{figure}

\section{Method}

Our novel method is based on the wavelet transform (WT), which is suitable for analyzing
timeseries containing non-stationary signal at different periods.
This property of the WT is due to its functions (wavelets), which are localized both in
time and frequency~\citep{1990ITIT...36..961D, 1997RvGeo..35..385K,1998BAMS...79...61T}.
The WT method provides amplitude and phase as well as power of the detected wave-pattern signals.
This approach has already been used for analysis of the solar radio signals in several articles
by different groups of authors ({\it e.g.} \opencite{2002ChJAA...2..183S, 2013A&A...555A..55Z,
2011A&A...525A..88M, 2011SoPh..273..393M, 2009ApJ...697L.108M}) and others.

The novelty of our approach is in the way that the essence of the WT results on non-stationary
wave-pattern signals is picked up and related to the time--frequency plots of the radio spectra.
The resulting parameters or each derived WT, {\it i.e.} amplitude, phase, and power, can be
presented separately within the time--frequency plots of the radio spectra.

Due to possible multiplicity of the non-stationary periodicities that can pre\-sent
simultaneously in the radio signals studied, one has to limit the investigated periodicities
either to that of maximum power in the general WT period range or to a narrower
WT period range of particular interest.
Both calculated significance of the derived WT periodicities and cone-of-influence can be
taken into account in our novel approach.

The method can be applied to any 2D time--frequency radio spectrogram.
The method is described in few consecutive steps given below:

\renewcommand{\labelenumi}{\roman{enumi})}
\begin{enumerate}
\item 
A linear trend of the radio signal, derived from the first and last values of the time series, is
subtracted separately from discrete data of all radio signal time series in the
2D time--frequency radio spectrogram. This operation keeps the first and last values of the series
close to zero.
\item
The WT of the adapted data of each individual radio-signal time series is performed
using the pre-selected mother wavelet type and its parameters (period (scale) range and
sampling, significance level resulting in the 2D WT amplitude, phase and power in
the time-period domain).
\item
The WT power spectra of all individual radio time series are checked for occurrence of
local power maxima in each temporal moment of the time series. The WT results
(power, amplitude, phase, significance) are taken together with the radio
frequency, time, and the WT period information for all such WT power maxima
locations detected. These local maxima have to be distinguished with a
clear decrease of the power to lower/higher periods.
\item
A histogram of the selected WT period results of the WT power local maxima above
a certain significance level is created from all individual radio time series
of the 2D time--frequency radio spectrogram.
This histogram is used for specifying a selected period range $[p_1,p_2]$ where a sufficiently
large number of period results are grouped in the period range of interest.
\item
The WT power local maxima results whose WT periods are localized within the selected period
ranges $[p_1,p_2]$ and their WT power is above the specified significance level are identified.
Such WT (amplitude, period, power) results are chosen for their time--frequency positions in
the 2D time--frequency radio spectrogram.
\item
These WT results can be then over-plotted on the original radio spectrum plots.
\end{enumerate}

The method described allows us to clearly identify frequencies where the
wave-pattern in the selected period range $[p_1,p_2]$ really exists, including its
temporal location and span. Moreover, this presentation of the identified frequencies/time/period
locations opens a possibility to select one of the resulting WT parameters to be plotted on top
of the radiospectrogram. We would like to point out that all WT calculations are performed
on the individual radio spectrum time series independently. Therefore a smooth behavior
of the selected WT resulting parameter over the identified frequency--time domain
with significant wave-pattern in the selected period range $[p_1,p_2]$ shows the reliability
of the derived WT results once the parameters of the WT are specified reasonably.

Finally we have to mention the actual parameters of the WT used in this study.
The Morlet mother wavelet, consisting of a complex sine wave modulated by a Gaussian,
was selected to search for radio signal variability at different periods.
The non-dimensional frequency $\omega_0$, equal to 6 was taken to satisfy the admissibility
condition~\citep{1992AnRFM..24..395F}. The WT has been calculated for the period range
from 10 to 600 seconds with scales sampled as fractional powers of two with ${\delta j = 0.4}$
still small enough to give an adequate sampling in scale, a minimum scale
${s_{j}=9.64}$\,s and a number of scales $N=101$.
As the basis of all calculations the WT computational algorithm
of \cite{1998BAMS...79...61T} was applied to individual radio-signal time series.

Due to the WT algorithm applied, the wavelet power spectra are suppressed within
a cone of influence \citep{1998BAMS...79...61T}. The cones are indicated in our
WT results plots by cross-hatched regions. Nevertheless, because we minimize the
radio signal at the beginning/end of the individual radio time series we take
into account all significant WT results, even those located within the aforementioned cones
of influence.

The significance level of the derived wavelet power spectrum was derived using the
null hypothesis according to \cite{1998BAMS...79...61T} assuming that the time series
has a mean power spectrum. If a wavelet power spectrum is well above the background
mean power spectrum, it can be assumed to be real with a certain confidence level.
The 95\,\% confidence level, used in this study, implies that 5\,\% of the wavelet power
should be above this level for each scale. The global wavelet spectrum was used as an
estimate of the background mean power spectrum against which the significance of the
local wavelet power spectrum features can be tested~\citep{1995Biometrika..82...619F,
1998JCli...11.2258K}.

The abovedescribed method finally allows us to identify and display the significantly
strong wave-pattern signals in the whole 2D time--frequency domain of the radio spectrogram.
The identified wave-pattern signals are of the power which peaks within the selected
period range for the particular temporal moment and investigated radio frequency.
Advantages of our novel method are that it provides a clear way to detect time--frequency
evolution of the radio signal wave-pattern appearance and at the same time to
follow the time--frequency distribution of the period, amplitude, and phase of those
wave-patterns. From the WT results over-plots on the original radio
spectrograms change of the periods, strength, and phase (frequency drift) of
the wave-pattern signals can be derived for interpretation of the radio source
properties.

\section{Results}

Using the abovedescribed method we analyzed the radio spectra presented in
Figure~\ref{fig1}. First, from radio-flux time series at all frequencies of
studied radio spectrum, we computed the wavelet power, wavelet amplitude, and
wavelet phase spectra. An example of such computations for the radio flux on
the 1600 MHz frequency, taken from the 800\,--\,2000 MHz Ond\v{r}ejov radio
spectrum, is shown in Figure~\ref{fig2}. These wavelet spectra show the period
of about 180\,seconds and periods above 300\,seconds.

Then, using such results for all frequencies in the selected radio spectrum, we
computed the histograms of all periods in this radio spectrum corresponding to
the well-defined local maxima of the WT power. Results for IZMIRAN, Phoenix-4,
and Ond\v{r}ejov radio spectra are shown in Figure~\ref{fig3}. The dotted
vertical lines indicate the periods 181, 69, and 40\,seconds, which were detected in UV
observations (\opencite{2011ApJ...736L..13L}). As seen here, while the
oscillation with the period of 181\,seconds was recognized in the whole 25\,--\,2000 MHz
radio spectrum, oscillations with the periods of 69 and 40 seconds were confirmed
only in the 250\,--\,870 MHz frequency range. In the 800\,--\,2000 MHz range we
found periods of 50 and 80\,seconds. Moreover, in the 250\,--\,870 MHz frequency
range, the oscillation with the period of about 420\,seconds was detected.

The most distinct period found both in the UV and radio observations is the
period 181\,seconds. Therefore, in Figures~\ref{fig4} and~\ref{fig5} we present
wavelet results computed for the period interval around this period, namely for
the interval 160\,--220\,seconds (the cross-hatched areas in Figure~\ref{fig3}).

Figure~\ref{fig4} shows the radio spectra recorded by the IZMIRAN, Phoenix-4,
and Ond\v{r}ejov radiospectrographs over-plotted by light-blue areas, where the
significant WT power for the period interval around 181\,seconds was found. On the
other hand, Figure~\ref{fig5} shows the radio spectra over-plotted by the areas
as in Figure~\ref{fig4}, but with the pink bands expressing the wave phase.
The black borders of these bands indicate zero phase. Note that at areas of
saturated data no periods and wave phases can be detected.

As seen in Figure~\ref{fig5} the phase of the 181\,second oscillation in this global
map appears to be nearly synchronized over the whole 25\,--\,2000 MHz frequency
range. However, a detailed view of some parts of the radio spectrum show a more
complex situation with drifts of the oscillation phase. For example, at the
beginning of the radio event in the time interval of 08:12\,--\,08:18 UT, the
zero oscillation phase drifts from the frequency 2000 MHz at 08:13:52\,UT, to
500\,MHz at 08:15:15\,UT (Figure~\ref{fig6}). The mean frequency drift of this
phase is about -18\,MHz s$^{-1}$. At lower frequencies, the phase of the
oscillations is more or less synchronized. For comparison see Figure~\ref{fig7},
showing the original radio spectrum in the same time and frequency intervals,.

Besides the most distinct period of 181\,seconds, found both in the UV and radio
observations, we have analyzed also the time--frequency localization of the less
dominant period of $\approx$69\,seconds. Here we can state that this period is highly
synchronized as in the case of the most dominant period in the period range 500\,--\,870 MHz.
The temporal span of the period occurrence is a little different.

\section{Analysis and Interpretation}

\begin{table}[]
\caption{Frequencies [$f$] and time difference [$\Delta t$] for the drifting zero
wave phase starting at 08:13:52\,UT  taken from Figure~\ref{fig6}, and
corresponding electron densities $n_{\rm e}$ (F-fundamental, H-harmonic) and
propagation distance [$D_{\rm w}$] of the UV wave with the velocity of $v_{\rm
w}$ = 2200 km s$^{-1}$.} \label{tab1}
\begin{tabular}{ccccc}
\hline
 ($f_1$ -- $f_2$)& $\Delta t (f_1-f_2)$  & $n_{\rm e} (f_1)$ & $n_{\rm e} (f_2)$  & $D_{\rm w} = v_{\rm w} \Delta t$ \\
$$[MHz]$$           &       [s]             & [cm$^{-3}$]      & [cm$^{-3}$]       & [km] \\
\hline
2000 -- 500 & 83 & 4.93 $\times$ 10$^{10}$ (F) & 3.08 $\times$ 10$^{9}$ (F) & 182,600  \\
2000 -- 500 & 83 & 1.23 $\times$ 10$^{10}$ (H) & 7.72 $\times$ 10$^{8}$ (H) & 182,600  \\
 \hline
\end{tabular}
\end{table}

\begin{table}[]
\caption{Heights [$h$] and distances [$D_{\rm Asch}$] according to the Aschwanden
model of the solar atmosphere for the plasma emission of the radio sources at
2000 and 500 MHz and time delays of propagation of the UV wave [$\Delta t_{\rm
w}$] ($v_{\rm w}$ = 2200 km s$^{-1}$) along these distances.}\label{tab2}
\begin{tabular}{cccc}
\hline
 $f$ [MHz]  &  $h$ [km]  & $D_{\rm Asch}$ [km] & $\Delta t_{\rm w}$ [s] \\
\hline
Fundamental \\
2000 &  8,250 &         &      \\
500  & 26,450 & 18,200  & 8.3  \\
\hline
Harmonic \\
2000 (1000) & 14,770 &        &      \\
  500 (250) & 47,370 & 32,600 & 14.8 \\
\hline
\end{tabular}
\end{table}

Now, let us analyze the phase drift, presented in Figure~\ref{fig6}, in more
detail. For this analysis we assume that the observed radio burst is generated
by the plasma emission mechanism, {\it i.e.} the electron plasma density can
be determined from the observed frequency as equal to the plasma or double
plasma frequency (emission on the fundamental (F) or harmonic (H) frequency).

Considering the above mentioned frequencies and times of the zero phase we
calculated the electron plasma densities in the radio sources at 2000 and 500
MHz. Then assuming that the phase drift is caused by the wave observed in UV
with the velocity $v_{\rm w}$ = 2200 km s$^{-1}$, we calculated the distance
[$D_{\rm w}$] of the UV wave propagation. The calculated plasma densities and
distances are shown in Table~\ref{tab1}.

Now, let us compare these distances with those in density models of the
solar atmosphere, {\it e.g.} models proposed
by~\citet{1947MNRAS.107..426A,1961ApJ...133..983N,1962ApJ...135..138M,1974SoPh...37..443P,1999A&A...348..614M,2002SSRv..101....1A}.
However, for the radio emission in the 500\,--\,2000 MHz range only the Aschwanden
model can be used. (Other  models we will consider in the analysis of the radio
emission at frequencies below 500 MHz.)

Thus, we compare the distances shown in Table~\ref{tab1} with those in
the density model by \citet{2002SSRv..101....1A}; the heights and distances are shown in
Table~\ref{tab2}, together with the propagation times of the UV wave ($v_{\rm w}$
= 2200 km s$^{-1}$) along these distances. As seen in
Tables~\ref{tab1} and~\ref{tab2}, the distance [$D_{\rm w}$] is much greater
than distances in the Aschwanden atmospheric model [$D_{\rm Asch}$]; compare also
$\Delta t (f_1-f_2)$ and $\Delta t_{\rm w}$. If we accept the idea that this
phase drift is caused by the wave observed in the UV, then there are two possible
explanations: i) the real density profile of the solar atmosphere differs
essentially from that in the Aschwanden model or ii) the angle between the UV wave
propagation and vertical direction in the solar atmosphere is large ($\approx$
80$^{\circ}$). Because the locations of radio sources are not known, we need to
consider both explanations, but based on the UV wave images it appears that the
first explanation is more probable. Note that combination of the two
explanations is also possible. Moreover, deviations from the Aschwanden
model may be due to density changes at the beginning of the flare.

\begin{table}[]
\caption{Distances [$D$] between heights, corresponding to 500 MHz and 25 MHz
plasma emission in the Aschwanden (Asch), 10$\times$ Baumbach-Allen (10BA), and
4$\times$ Newkirk (4N) models of the solar atmosphere. Times [$t$] mean those for
propagation of the electron beam with the velocity c$/3$.}\label{tab3}
\begin{tabular}{cccccc}
\hline
 $D_{\rm Asch}$ & $t_{\rm Asch}$ &  $D_{\rm 10BA}$  & $t_{\rm 10BA}$ & $D_{\rm 4N}$ & $t_{\rm 4N}$ \\
$$[km]$$   & [s] & [km] & [s]  & [km] & [s]   \\
\hline
Fundamental \\
 255,600 & 2.55 & 965,400 & 9.65 & 1,103,300 &  11.03 \\
\hline
Harmonic   \\
 330,400 & 3.30 & 1,298,300 & 12.98 & 2,015,200 & 20.15 \\
\hline
\end{tabular}
\end{table}

Furthermore, Figure~\ref{fig6} shows that in the 25\,--\,500 MHz frequency
range the oscillation phase is more or less synchronized. The corresponding
distances in the Aschwanden, 10$\times$ Baumbach--Allen (10BA), and 4$\times$
Newkirk models are shown in Table~\ref{tab3}. The UV wave is too slow to
explain this synchronization at such large distances. Thus, it appears that only
the possible explanation of this synchronization is the electron beams, whose
acceleration is modulated by the UV wave, somewhere in the deep and dense layers of
the solar atmosphere. Owing to this modulation, the electron beams are
accelerated with the period of the UV wave (181 seconds). These beams propagate
upwards through the solar corona and generate the 25\,--\,500 MHz radio emission
with the 181 seconds period. Due to high beam velocity, the 25\,--\,500 MHz radio emission,
corresponding to a large interval of heights in the solar corona, is nearly
synchronized. If we assume the beam velocity as $\approx$c$/3$ = 100,000 km
s$^{-1}$, where c is the light speed, then the Aschwanden model is the best
of the models considered for explanation of the 25\,--\,500 MHz synchronization within
a few seconds; see the times in Table~\ref{tab3}. This explanation can be supported
by Type III bursts observed in the 25\,--\,270 MHz range in the radio spectrum, as shown in
Figure~\ref{fig7}.

As concerns other parts of the phase map (Figure~\ref{fig5}), some
drifts at higher frequencies can be explained by the propagation of the UV wave
and synchronization of the oscillation phase at lower frequencies (higher
altitudes in the solar atmosphere) by Type III electron beams.

Finally, we estimated the velocity of the agent generating the maximum burst
drift (BM), observed in the 08:22\,--\,08:50 UT interval. Using all of above
mentioned models, the velocity was estimated as 6\,--\,46 km s$^{-1}$. Because
this velocity is too slow, it appears that it expresses a growth of the flaring
magnetic-field structure during this time interval.

\section{Conclusions}

In order to search for radio signatures of the UV waves that were observed
in the 1 August 2010 event, we constructed a new type of map using the wavelet
technique applied on the broadband 25\,--\,2000 MHz radio spectrum.

An analysis of UV observations during the 1 August 2010 event revealed
periods of 181, 69, and 40 seconds. In the present article, in the radio range, we found
the same periods: i) The oscillation with the period of 181 seconds was found in the
whole 25\,--\,2000 MHz radio spectrum, and ii) the oscillations with the periods
of 69 and 40 seconds were only in the 250\,--\,870 MHz range.  Moreover, in the
800\,--\,2000 MHz frequency range we found periods of 50 and 80 seconds, and in the
250\,--\,870 MHz range, an oscillation with a 420-second period.

We made maps of phases of the 181 second oscillations in order to analyze
their frequency drift. Namely, assuming that the plasma emission mechanism
generated the radio bursts in the 1 August 2010 event, the drift of the
oscillation phase can be interpreted as caused by the UV wave propagating
upwards (negative frequency drift) or downwards (positive frequency drift)
through the solar atmosphere.

We found that at the beginning of the radio event, in the 2000\,--\,500 MHz
frequency range, the phase of the 181-second oscillation drifts towards lower
frequencies. This drift we interpreted as caused by the propagating UV wave.
However, because the travel distance of the UV wave with the velocity $v_{\rm
w}$ = 2200 km s$^{-1}$ is much greater than the height differences
(corresponding to the 2000\,--\,500 MHz frequencies) in the model of the solar
atmosphere by \cite{2002SSRv..101....1A}, we propose that in this case the real
electron=density profiles differs from that in the Aschwanden model or the
propagation direction of the UV wave deviates substantially from vertical
direction in the solar atmosphere. Based on the UV wave images, it appears that
the first explanation is more probable. Note that combination of the both
explanations is also possible. Moreover, deviations from the Aschwanden
model may be due to density changes at the beginning of the flare.

On the other hand, at frequencies below 500 MHz we found that the phase is more
or less synchronized (within a few seconds). We propose that this phase
synchronization is caused by electron beams, of which the acceleration is
modulated by the UV wave, somewhere in the deep and dense layers of the solar
atmosphere. Owing to this modulation, the electron beams are accelerated with
the period of the UV wave (181 seconds). These beams propagate upwards through
the solar corona and generate the 25\,--\,500 MHz radio emission with the
181-second period. Due to high beam velocity, the 25\,--\,500 MHz radio
emission, corresponding to a large range of heights in the solar corona, is
nearly synchronized. If we assume the beam velocity as $\approx$c$/3$, where c
is the light speed, then the Aschwanden model is the best of the models
considered for an explanation of the 25\,--\,500 MHz synchronization within a
few seconds. Type III bursts observed in this range speak in favour of this
interpretation.

Finally, while the processes of modulation of the electron beam acceleration by
magnetosonic waves were considered in several articles,{\it e.g.}, by
\cite{2016SSRv..200...75N}, the process of generation or modulation of the
plasma radio emission by the UV wave is unclear. Maybe, at some locations the
propagating UV (fast magnetosonic) wave is changed to a weak magnetosonic shock
or accompanied by it. The shock accelerates electrons, of which the
distribution function is unstable for plasma waves producing radio emission,
similarly to that in Type II burst. Another possibility is that the density
variations modify the optical thickness of the radio emission, especially for
the emission close to the plasma frequency.

\begin{acks} 
The authors thank an anonymous referee for useful comments.
Furthermore, the authors thank V. Fomichev and R. Gorgutsa for providing
of the IZMIRAN data. This research was supported by the grants P209/12/0103 (GA
\v{C}R) and the project RVO:67985815 of the Astronomical Institute of the
Academy of Sciences.
This work was supported by the Science Grant Agency project VEGA
2/0004/16 (Slovakia).
Help of the Bilateral Mobility Project SAV-16-03 of the SAS and CAS is acknowledged.
This article was created by the realization of the project ITMS No.26220120009, based on the
supporting operational Research and development program financed from the European Regional
Development Fund.
This research has made use of NASA's Astrophysics Data System.
The wavelet analysis was performed with software based on tools provided by
C. Torrence and G. P. Compo at {\tt paos.colorado.edu/research/wavelets}.
\end{acks}

\section*{Disclosure of Potential Conflicts of Interest}

The authors declare that they have no conflicts of interest.

\bibliographystyle{spr-mp-sola}
\bibliography{ref}

\end{article}
\end{document}